\documentclass[aps,prstab,twocolumn,superscriptaddress,a4paper,amsfonts,amssymb,amsmath,showpacs]{revtex4-1}

\usepackage{lipsum}
\usepackage{bm}
\usepackage{pstool}
\usepackage[caption=false]{subfig}
\usepackage{tikz}
\usetikzlibrary{arrows,calc,shapes}
\usepackage{amsmath,amsthm,amssymb}
\usepackage{relsize}
\usepackage{mathrsfs}
\usepackage{wasysym}

\usepackage{graphicx,picture,calc}
\usepackage[urlcolor=blue,breaklinks=true]{hyperref}
\hypersetup{colorlinks, linkcolor=blue, urlcolor=blue, citecolor=blue}
\allowdisplaybreaks

\usepackage[colorinlistoftodos]{todonotes}
\usepackage{siunitx}
\usepackage{csquotes}
\usepackage{placeins}

\newcolumntype{L}[1]{>{\raggedright\let\newline\\\arraybackslash\hspace{0pt}}m{#1}}
\newcolumntype{C}[1]{>{\centering\let\newline\\\arraybackslash\hspace{0pt}}m{#1}}
\newcolumntype{R}[1]{>{\raggedleft\let\newline\\\arraybackslash\hspace{0pt}}m{#1}}

\begin{document}
\title{Identification and characterization of high order incoherent space charge driven \\ structure resonances in the CERN Proton Synchrotron}

\author{F.~Asvesta}
\email[]{foteini.asvesta@cern.ch}
\affiliation{CERN, CH 1211 Geneva 23, Switzerland}
\affiliation{National Technical University of Athens,
15780 Zografou, Greece}

\author{H.~Bartosik}
\affiliation{CERN, CH 1211 Geneva 23, Switzerland}

\author{S.~Gilardoni}
\affiliation{CERN, CH 1211 Geneva 23, Switzerland}

\author{A.~Huschauer}
\affiliation{CERN, CH 1211 Geneva 23, Switzerland}

\author{S.~Machida}
\affiliation{STFC Rutherford Appleton Laboratory, OX11 0QX, Didcot, United Kingdom}

\author{Y.~Papaphilippou}
\affiliation{CERN, CH 1211 Geneva 23, Switzerland}

\author{R.~Wasef}
\email[]{raymond.wasef@hec.edu}
\affiliation{CERN, CH 1211 Geneva 23, Switzerland}

\date{\today}

\begin{abstract}

Space charge is typically one of the performance limitations for the operation of high intensity and high brightness beams in circular accelerators.   
In the Proton Synchrotron (PS) at CERN, losses are observed for vertical tunes above $Q_y=6.25$, especially for beams with large space charge tune shift. 
The work presented here shows that this behaviour is associated to structure resonances excited by space charge due to the highly symmetric accelerator lattice of the PS, typical for first generation alternating gradient synchrotrons. 
Experimental studies demonstrate the dependency of the losses on the beam brightness and the harmonic of the resonance, 
and simulation studies 
reveal
the incoherent nature of the resonance.
Furthermore, the calculation of the Resonance Driving Terms (RDT) generated by the space charge potential shows that the operational working point of the PS is surrounded by multiple space charge driven incoherent resonances.
Finally, measurements and simulations on both lattice driven and space charge driven resonances
illustrate the different behaviour of the beam loss depending on the source of the resonance excitation and on the beam brightness.
\end{abstract}

\maketitle

\section{Introduction}

Space charge effects in combination with betatron resonances can be one of the main performance limitations for high brightness beams in circular accelerators.
The space charge force generates an incoherent tune shift that depends on the line density of the longitudinal beam profile and the transverse beam size evolution around the machine. 
Different tune shifts of individual particles lead to a tune spread in the transverse tune space.
The interplay between the space charge tune spread and excited resonances 
was previously studied on a controlled normal octupole resonance~\cite{ref:Franchetti2003,ref:Metral2006, ref:Franchetti2006}. The experimentally observed transverse emittance growth and beam loss was identified to be caused by the periodic resonance crossing and trapping of individual particles due to the modulation of the transverse space charge force through the synchrotron oscillations.
More recently, the same mechanism was studied at a coupled sextupole resonance~\cite{ref:Franchetti2017}.
However, in addition to the induced incoherent tune spread, the non-linear transverse space charge potential can also directly excite structure resonances, as shown in~\cite{MACHIDA1997316, ref:Cousineau2003, Lee_2006}.

\begin{figure}[tb]
    \centering
        \includegraphics[width=0.95\columnwidth]{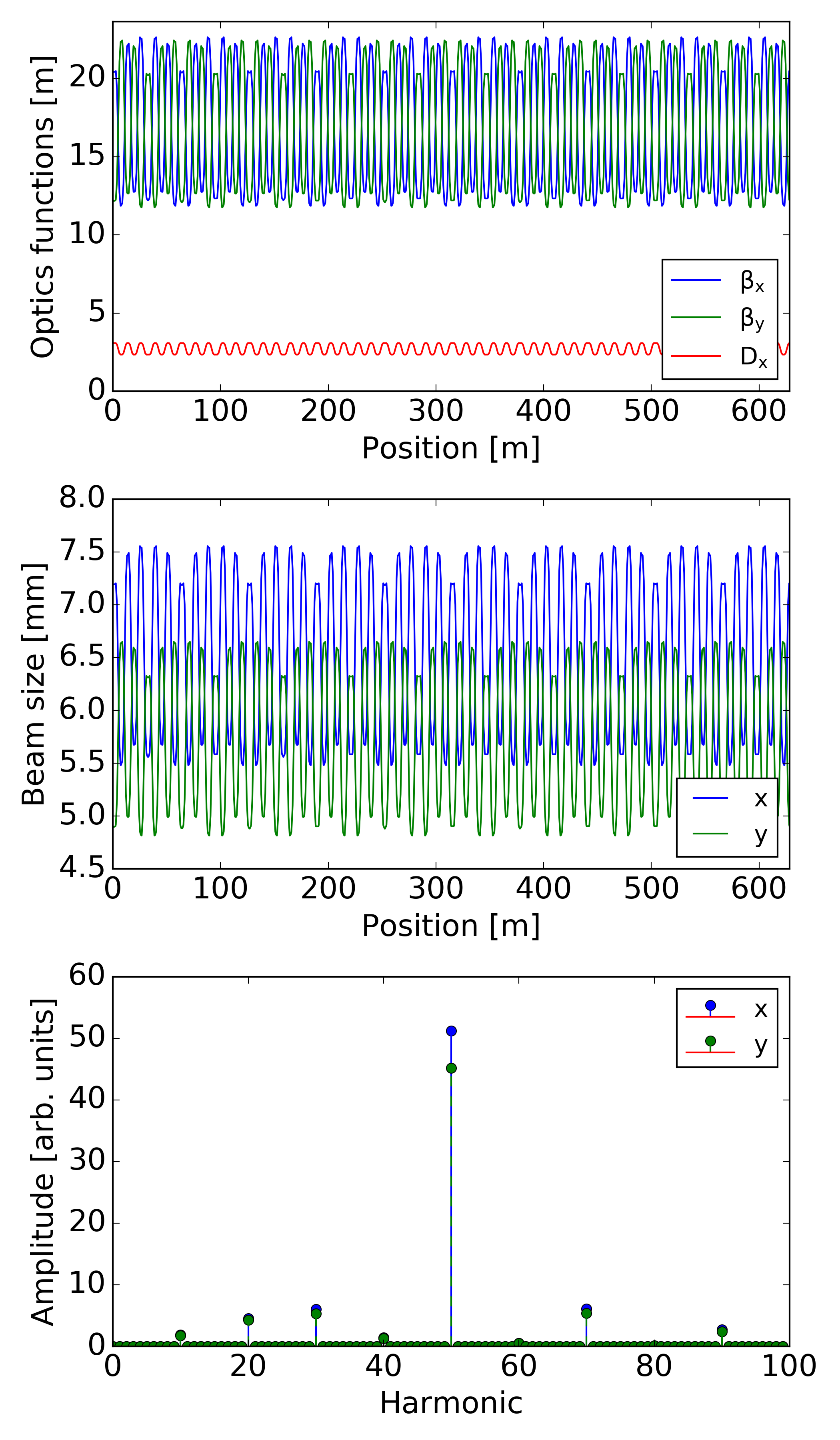}
        \caption{Linear optics functions in the PS lattice (top) and the corresponding beam size (center) calculated using transverse normalized 1 $\sigma$ emittances $\varepsilon_x^n=5.5$, $\varepsilon_y^n=4.5$ and an rms momentum spread of $\Delta p/p=0.52\times10^{-3}$. Harmonic analysis of the beam size modulation (bottom).}
    \label{fig:psbeamSize}
\end{figure}

The CERN Proton Synchrotron (PS) provides an excellent test bench for the study of space charge effects due to its highly symmetric lattice and the long beam storage time required at low energy in normal operation~\cite{Damerau:IPAC2018-WEPAF063}.
Moreover, space charge is the major performance limitation for high brightness beams required in the context of the LHC Injectors Upgrade (LIU) project~\cite{ref:LIU}.
The PS consists of 100 combined function main magnets, each divided into a focusing and defocusing half unit,
resulting in a total of 50 cells. 
The bare tunes of $Q_x=6.25$, $Q_y=6.28$ are determined by the quadrupolar components of the main magnets.
The tunes can be adjusted either using dedicated quadrupoles or circuits of pole face windings (PFW) on the main magnet poles~\cite{ref:WorkingPoint}. Each of these two options comes with disadvantages, namely a large perturbation of the periodicity of the machine for large tune adjustments with the dedicated quadrupoles (as they are not placed symmetrically around the ring), while the tune control through the PFWs also generates unavoidable higher order field components.
In the usual operation of the PS, the dedicated quadrupoles are preferred for the tune adjustment at low energy, while the PFW are used to control tunes and chromaticities during the ramp.
Even though the 50-fold periodicity of the bare lattice is slightly perturbed by the insertion of 20 long straight sections, the optics remain highly regular and the main harmonic of the lattice is 50, as can be seen through the optics and the harmonic analysis of the beam size in Fig.~\ref{fig:psbeamSize}. 
The presence of this strong lattice harmonic suggests that resonances of the same harmonic, i.e. $mQ_x+nQ_y=50$, where $(m,n)$ are integer numbers indicating the order of the resonance, 
can be driven by systematic errors in the machine and even by space charge itself.

The maximum incoherent space charge tune shifts for the operational LHC beams at injection in the PS are $\Delta Q_x\approx-0.19$ and $\Delta Q_y\approx-0.24$ and the bare tunes are usually set to $Q_x=6.20$ and $Q_y=6.24$.
The need to accommodate an even higher incoherent space charge tune shift of beams with even higher brightness in the course of the LIU project drives the need for detailed resonance and space charge studies in order not to exceed the allocated 5~\% emittance growth and 5~\% loss budgets.

\begin{figure}[!t]
	\includegraphics[width=0.99\columnwidth]{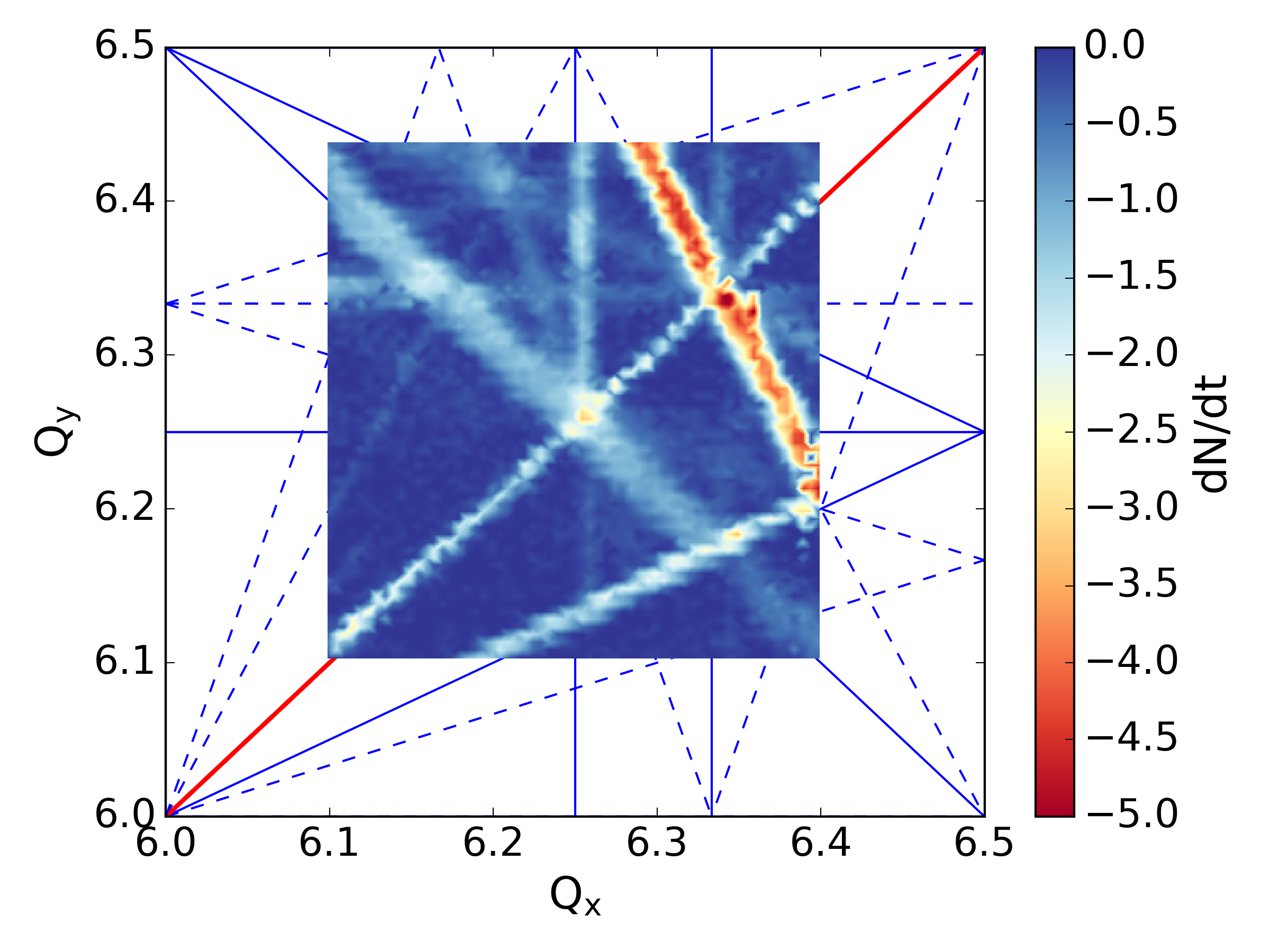}
	\caption{Beam loss map in the transverse tune space 
	colour coded to the loss rate. Theoretical resonance lines up to $\mathrm{4}^{\mathrm{th}}$ order are plotted, systematic ones in red and non-systematic ones in blue.
	The skew resonances are shown with dashed lines and the normal resonances with solid lines.}
	\label{fig:alex}
\end{figure}

To guide the choice of the optimal working point for these high brightness beams, dynamic tune scan measurements were performed using a low brightness beam with large transverse emittances in order to identify excited resonances through the recorded losses, as reported in the past for the PS~\cite{Huschauer:1501943} and the CERN Proton Synchrotron Booster (PSB)~\cite{Santamaria_Garcia:IPAC19-MOPTS086}.
During the measurements one of the tunes is kept constant while the other is varied linearly to cover the available tune space.
The procedure is repeated until the accessible tune space is fully covered, in all possible directions. 
Measuring the beam loss rate as a function of the tune settings allows the identification of excited resonances as shown in Fig.~\ref{fig:alex}. 

\begin{figure}[!t]
	\includegraphics[width=0.95\columnwidth]{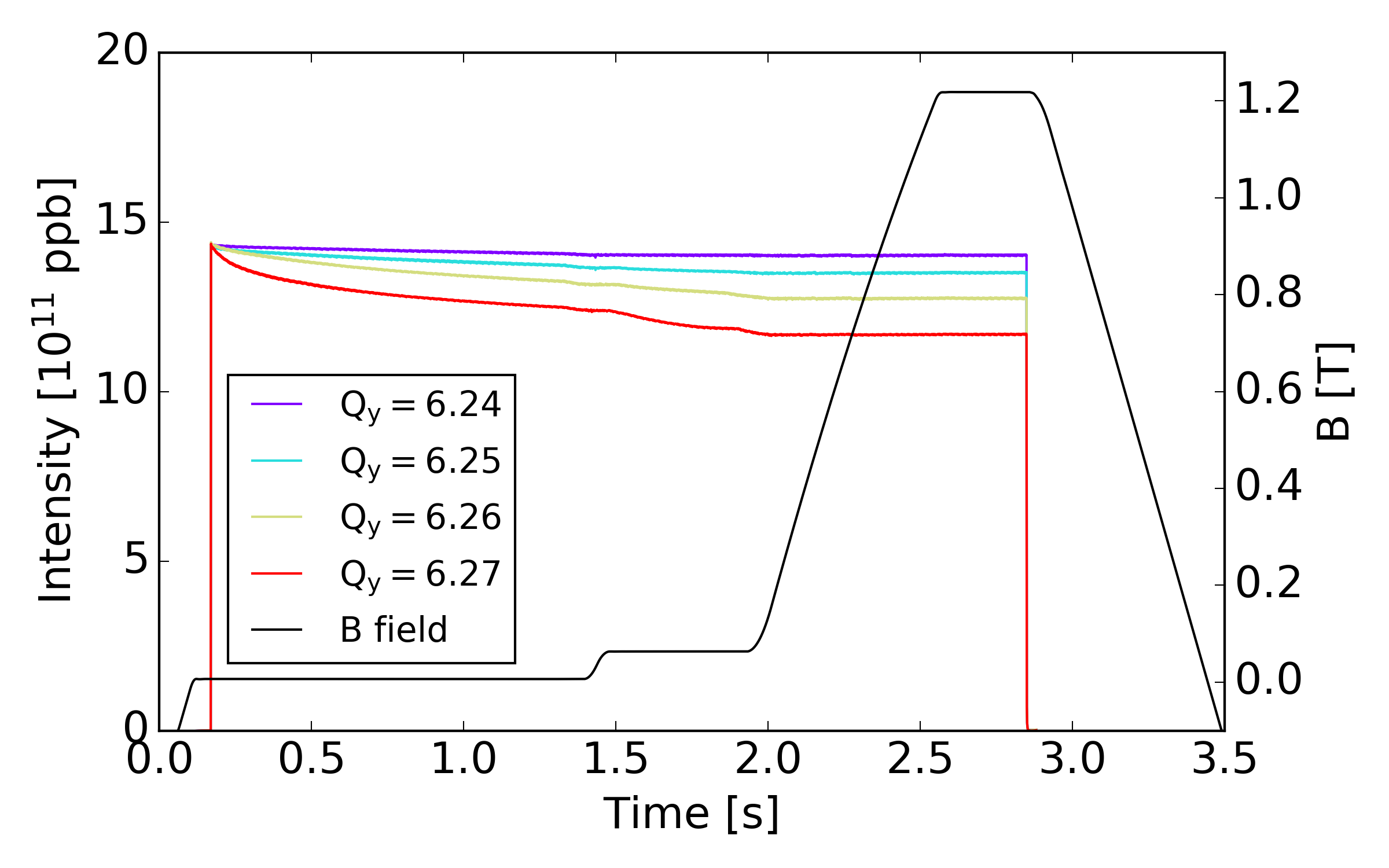}
	\caption{Intensity along the PS cycle for different vertical tunes. The magnetic field along the cycle is plotted on a second axis.}
	\label{fig:intensity}
\end{figure}

Various low order resonances up to $\mathrm{4^{th}}$ order can be clearly identified. In particular, the strongest resonance seems to be the sum resonance $2Q_x+Q_y=19$, most likely excited by skew sextupole errors in the main magnets~\cite{Schoerling:IPAC14-TUPRO107}. 
However, since the exact error remains unknown no skew sextupole-like errors are included in the nonlinear PS optics model.
Other sextupole resonances such as the $3Q_y=19$ skew resonance as well as the $Q_x-2Q_y=-6$ normal sextupole resonance appear considerably weaker. 
Furthermore, the excitation of the resonances $2Q_x+2Q_y=25$ and $4Q_x=25$ indicates the presence of octupolar-like errors.
On the other hand, the main region of interest around the operational working point $Q_x=6.20,~Q_y=6.24$ seems free of resonances when considering the results from dynamic tune scans. 
Therefore, one could expect that working points above $Q_y=6.25$ could be suitable for operating high brightness beams in the PS. 
However, experimental studies to explore higher vertical tunes showed slow losses in the low energy part of the PS cycle as presented in Fig~\ref{fig:intensity}. 
It could be argued that the identification of resonances through dynamic tune scans is not sensitive enough to resolve relatively weak resonances. Instead, it was shown that the nonlinear space charge potential excites a structure resonance at $Q_y=6.25$ due to the high periodicity of the PS  lattice~\cite{Gilardoni:2131161,RayPhD}. As shown in the following sections, a detailed analysis of the space charge driven structure resonances in the PS reveals that the operational working point is actually surrounded by several space charge driven $8^\textrm{th}$ order incoherent structure resonances, as identified through experimental observations combined with simulations and analytical studies. 

This paper is organized as follows. In Section~\ref{sec:resonance-identification}, 
the initial studies leading to the identification of a space charge driven structure resonance at $Q_y=6.25$ are described.
A detailed characterization through simulation studies is presented in Section~\ref{sec:resonance-characterization}, clearly highlighting the incoherent nature of this resonance. 
With this result, the Hamiltonian resonance driving terms induced by the space charge potential of a Gaussian beam are calculated analytically in Section~\ref{sec:RDTs}, revealing additional space charge driven structure resonances. 
The resonance excitation is also demonstrated using the Frequency Map Analysis (FMA) technique in Section~\ref{sec:fma}.  
The expected loss mechanism for bunched beams is discussed in Section~\ref{sec:lossmechanism}. Section~\ref{sec:machinestudies}
summarizes detailed experimental and simulation studies performed with low and high brightness beams, confirming the presence of various space charge driven structure resonances in the PS. 
Conclusions are given in Section~\ref{sec:conclusions}.

\section{Experimental Identification Of Structure Resonances}\label{sec:resonance-identification}


The resonance at $Q_y=6.25$ leading to the losses observed in Fig.~\ref{fig:intensity} could be driven either by random octupole-like errors of the PS lattice, or by the space charge potential due to the structure of the lattice.
To experimentally identify the source of the excitation a couple of experiments were designed in order to correlate the observations to brightness and the $50^\textrm{th}$ lattice harmonic. 

\begin{figure}[!b]
    \centering
        \includegraphics[width=0.95\columnwidth]{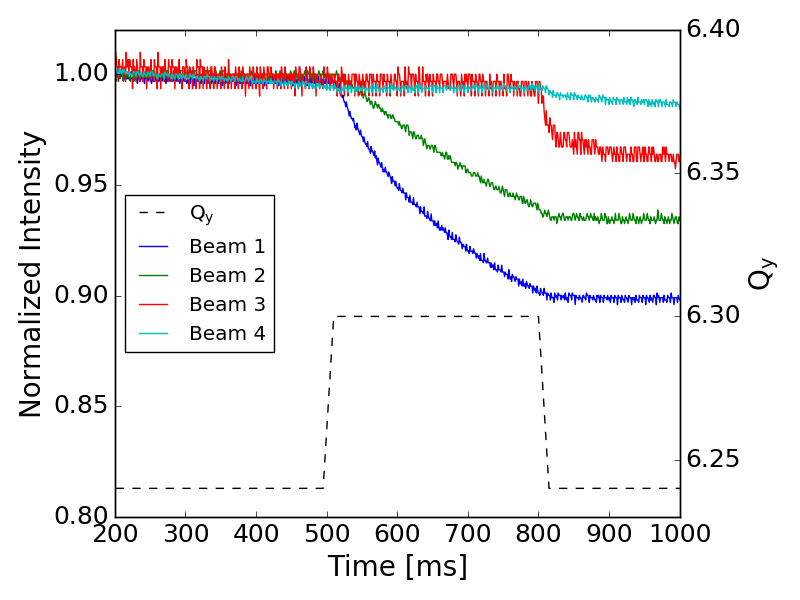}
        \caption{Normalized intensity of four beams with different space charge tune shifts when the $Q_y=6.25$ resonance is crossed (solid colour lines). The dashed line shows the vertical tune step on a second axis.}
    \label{fig:loss_varying_brightness}
\end{figure}

The dependence of the losses on the strength of the direct space charge force is demonstrated by an experimental study with beams of varying brightness.
The beam parameters are summarized in Table~\ref{tab:beams},
the intensity is given in particles per bunch (ppb), the momentum deviation is noted as $\Delta p/p$ and the normalized transverse emittances as $\varepsilon_{\mathrm{x,y}}^{\mathrm{n}}$. 
During the experiment, the horizontal tune was kept at 6.23 and the vertical tune was ramped from 6.24 to 6.30 and it was kept constant for 300\,ms, before ramping it down again to 6.24. 
The advantage of such a tune step is that beam loss starts and ends at precise times and can be clearly associated to the crossing of the resonance through the increase of the vertical tune. Figure~\ref{fig:loss_varying_brightness} shows the relative intensity of the four beams along the cycle and the evolution of the vertical tune.
It can be seen that the beam loss is directly correlated to the space charge force, demonstrated by the fact that the beam type 1 (blue) with the highest tune shift experiences the highest losses. 

\begin{table}[!t]
 \caption{Beam parameters for the experimental results shown in Fig.~\ref{fig:loss_varying_brightness}}
  \centering
  \begin{tabular}{L{3.2cm} C{1.1cm} C{1.1cm} C{1.1cm} C{1.3cm}}
  \toprule
       Beam ID & 1 & 2 & 3 & 4 \\
       \colrule
       Intensity [\SI{e10}{ppb}]  & 115 & 80 & 35 & 115 \\
       $E_\textbf{kin} [\SI{}{GeV}]$ & 1.4 & 1.4 & 1.4 & 1.4\\
       Bunch length ($4\sigma$)~[\SI{}{ns}] & 100 & 100 & 100 & coasting\\ 
       $\Delta p/p~(\textrm{rms})~[10^{-3}]$& 1.8 & 1.8 &1.8& 1.8\\ 
       $\varepsilon_{\mathrm{x}}^{\mathrm{n}}$~($1\sigma$)~[\SI{}{\micro\meter}] & \SI{1.3}{} & \SI{0.77}{}& \SI{0.95}{}& \SI{1.3}{}\\ 
       $\varepsilon_{\mathrm{y}}^{\mathrm{n}}$~($1\sigma$)~[\SI{}{\micro\meter}] & \SI{1.6} & \SI{1.1}{} & \SI{0.6}{} & \SI{1.6}{} \\ 
       $\Delta Q_x$ (maximum) & -0.22 & -0.18 & -0.08 & -0.01 \\
       $\Delta Q_y$ (maximum) & -0.40 & -0.37 & -0.24 & -0.01 \\  
      \colrule
  \toprule
  \end{tabular}
 \label{tab:beams}
\end{table}

To verify that the resonance is only excited at the $\mathrm{50^{th}}$ harmonic, an experiment at different integer tunes was performed to probe the excitation at other resonance harmonics. 
The change of the vertical tune by one integer was achieved using an extra circuit installed in the main magnets referred to as Figure-of-8-Loop (F8L), which has the advantage of changing the quadrupolar field component of the combined function magnets without significantly exciting higher order field components~\cite{ref:Mariusz}.
The main limitation of this circuit is that it acts on both transverse tunes in opposite directions, as it increase the field in one half unit while decreasing it in the other.
Hence, the experiment was conducted with the horizontal tune decreased by one integer unit and the vertical tune increased by one unit, which is referred to as $(5,7)$ optics according to the integer parts of $(Q_x,Q_y)$.
It was verified that the integer part of the tune was indeed 5 in the horizontal and 7 in the vertical plane. 
The measured optics agreed with the model within 10\% beta-beating and dispersion-beating~\cite{RayPhD}.

\begin{figure}[tb]
  \includegraphics[trim=10 0 0 0, clip, width=0.86\columnwidth]{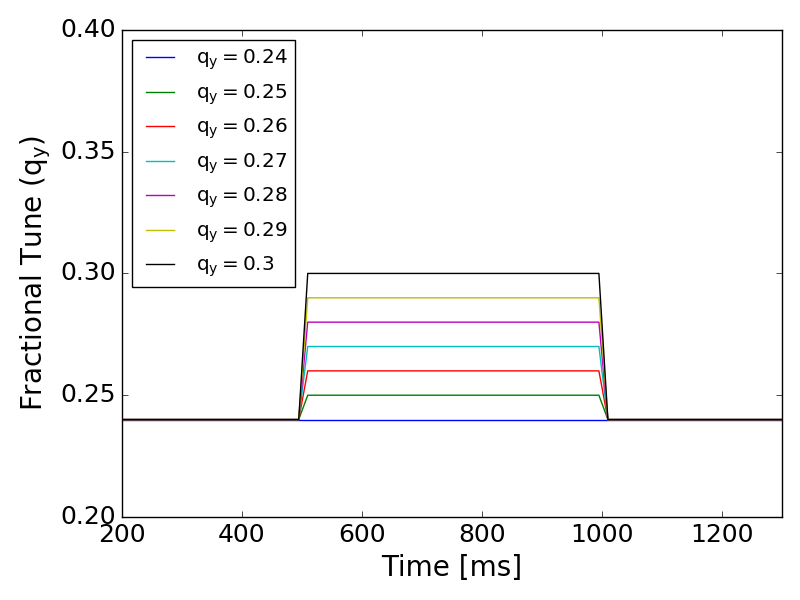}
  \includegraphics[width=0.86\columnwidth]{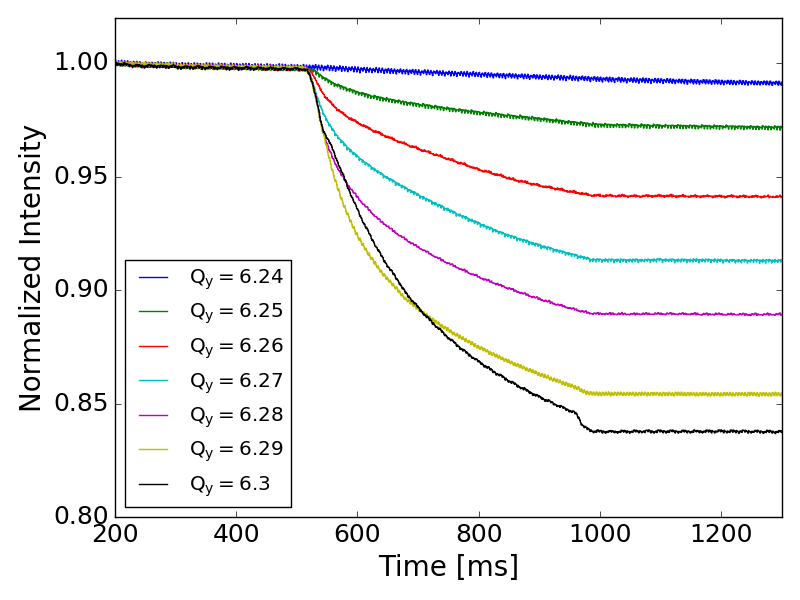}
  \includegraphics[width=0.86\columnwidth]{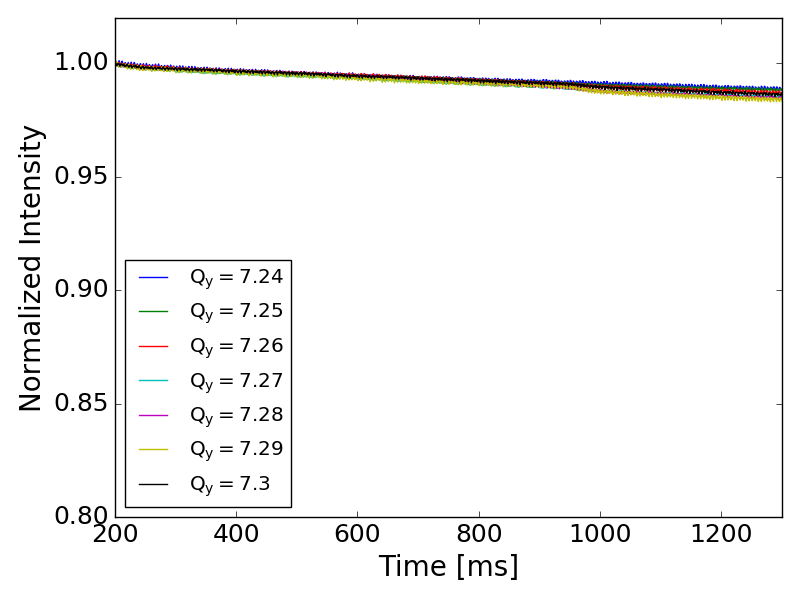}
\caption{Vertical tune steps programmed in the experiments (top).  Normalized intensity along the cycle for the (6,6) nominal optics (center) and the (5,7) split-tunes optics (bottom).}
\label{fig:tunescanRay}
\end{figure}

The main measurement consisted of a tune-step in which the horizontal tune was kept constant while the fractional vertical tune was changed from 0.24 to a plateau value, where it was kept constant for 500\,ms and then changed back to 0.24. The plateau value was varied between 0.24 and 0.3 in order to see the effect at the different working points, as shown in Fig.~\ref{fig:tunescanRay} (top). This measurement was performed in the (6,6) nominal optics and the (5,7) split-tunes optics. In the case of the nominal optics shown in Fig.~\ref{fig:tunescanRay} (center), the beam loss clearly increases for higher working points (since more particles cross the resonance due to the space charge induced tune shift). 
On the contrary, no beam loss was observed for the same range of fractional tunes in the case of the split-tunes (5,7) optics, as shown in Fig.~\ref{fig:tunescanRay} (bottom).
The change of integer was found to be effective in mitigating against harmful effects of the resonance, confirming that the resonance at $Q_y=6.25$ is a space charge driven structure resonance. 
Unfortunately, the split-tunes optics (5,7) cannot be used in routine operation, as the available strength of the F8L is not sufficient to maintain this working point up to extraction energy. In addition, the impact on the gamma transition jump scheme~\cite{Burnet:1359959} due to the change of phase advance between the special fast quadrupoles would require further studies.

\section{Resonance characterization}
\label{sec:resonance-characterization}

The presence of the strong $\mathrm{50^{th}}$ harmonic in the lattice and the excitation of the resonance at $Q_y=6.25$ through space charge was experimentally confirmed.
However, the resonance could be excited either in $\mathrm{8^{th}}$ order as an incoherent structure resonance $8Q_y=50$, or in $\mathrm{4^{th}}$ order, as a coherent parametric resonance $4Q_y=50/2$~\cite{Okamoto:2002zm}.
The order of the excitation and hence the nature of the resonance at $Q_y=6.25$ can be verified in simulation studies.
For this purpose, the resonance was dynamically crossed~\cite{Hofmann:hb2016, Hofmann:2287897} using a coasting beam.
In particular, the simulations are performed with two different space charge models: 1) self-consistent space charge solvers, which take into account both the coherent and incoherent response of the beam, 2) frozen model without update using analytical solvers for Gaussian distributions based on the Bassetti-Erskine formula~\cite{Bassetti:1980by}, which  take into account incoherent effects only. Comparing the results obtained with these two simulation models allows to identify the nature of the resonance. 
Furthermore, long term tracking simulations are used to study the agreement of these two models for bunched beams and long storage times.

\subsection{Simple FODO lattice (coasting beam)}

\begin{figure}[tb]
    \centering
        \includegraphics[width=0.95\columnwidth]{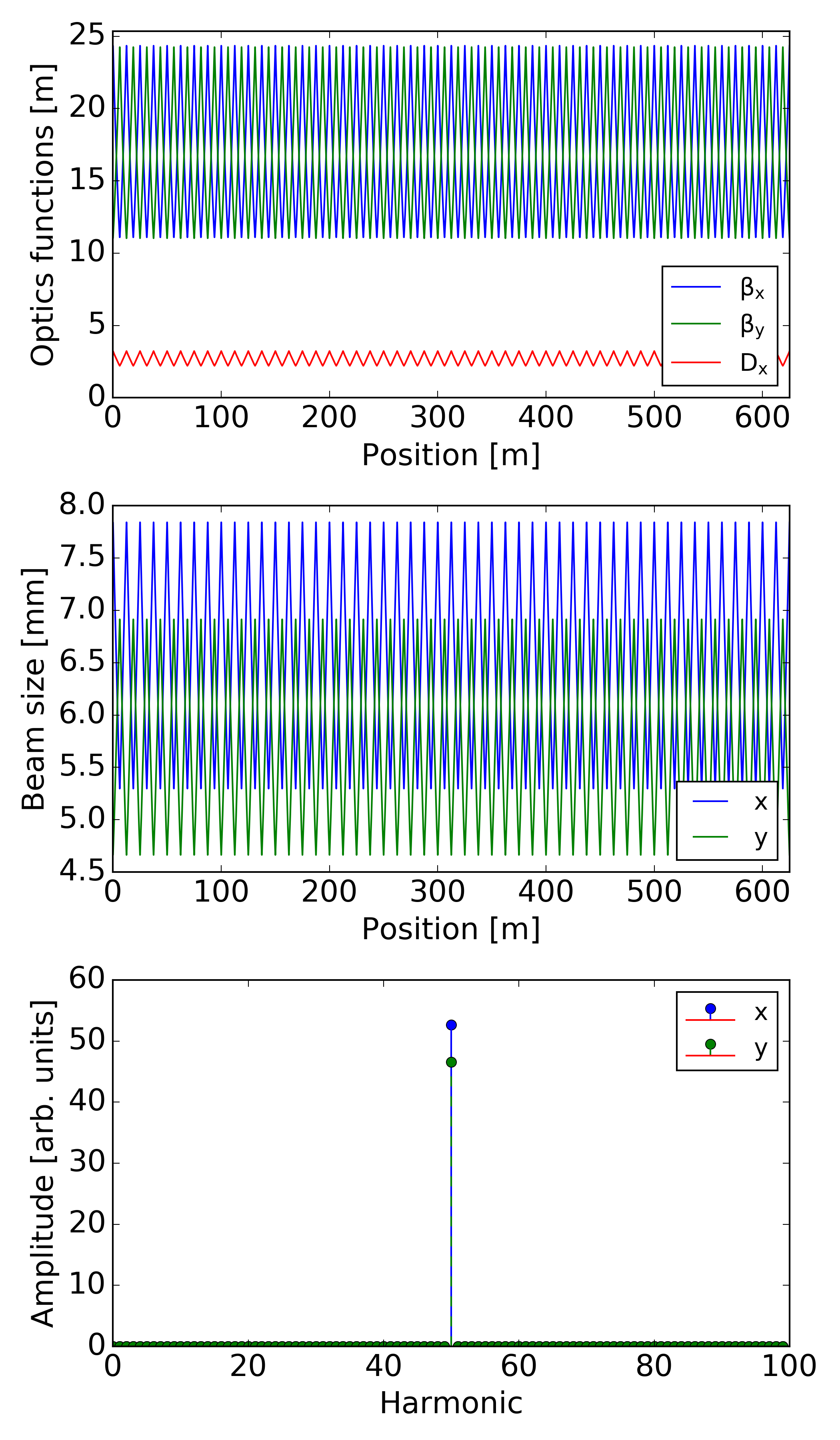}
        \caption{Linear optics functions in the simple FODO lattice (top) and the corresponding beam size (center) calculated using transverse normalized 1 $\sigma$ emittances $\varepsilon_x^n=5.5$, $\varepsilon_y^n=4.5$ and an rms momentum spread of $\Delta p/p=0.52\times10^{-3}$. Harmonic analysis of the beam size modulation (bottom).}
    \label{fig:fodoOptics}
\end{figure}

Initially a simplified FODO lattice is used in order to study space charge driven resonances, like the one observed at $Q_y=6.25$ in the PS, without taking into account the full complexity of the PS lattice yet.
The FODO lattice used here consists of 50 identical cells, which is similar to the PS lattice, but without the 20 slightly longer straight sections.
It has the same periodicity, energy and circumference, and therefore it can be matched to the same working points and excite the same structural resonances.
The optics functions, the beam size evolution and the lattice harmonics of the FODO lattice are shown in Fig.~\ref{fig:fodoOptics}.
The resonance at $Q_y=6.25$ is dynamically crossed by varying the tune in the vertical plane, while the horizontal tune is kept constant at $Q_x=7.2$. 
Note that the horizontal tune is moved to a different integer in order to avoid the crossing of the extra space charge driven resonances that will be discussed in Section~\ref{sec:RDTs}.
The lattice is matched in MAD-X~\cite{MADX} and the tracking simulation is performed with  PTC~\cite{Schmidt:573082} in PyOrbit~\cite{PyORBIT}. 
A coasting beam is used in this study in order to avoid the complexity coming from the longitudinal motion, and to enhance any possible coherent response of the beam.
The beam is generated with Gaussian transverse distributions. 
The space charge force is included in the simulations using either the fully self-consistent 2.5\,D Particle-In-Cell (PIC) solver, in which the force is calculated using a Poisson solver on a grid and weighted using the longitudinal line density, or an analytic solver in which the space charge force is evaluated based on a frozen potential calculated from the lattice functions and the macroscopic parameters of the beam using the Bassetti-Erskine formula~\cite{Bassetti:1980by}. 
The beam parameters are chosen to obtain maximum tune shifts in the order of $\Delta Q_x\approx-0.24$ and $\Delta Q_y\approx-0.34$.

\begin{figure*}[p] 
    \centering
        \includegraphics[width=0.45\textwidth]{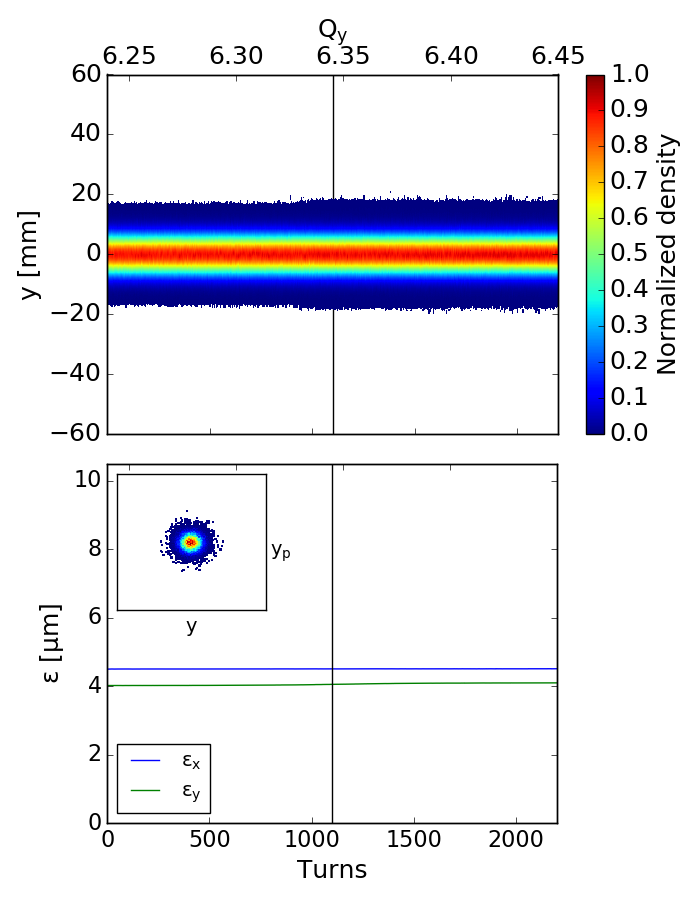}
        \includegraphics[width=0.45\textwidth]{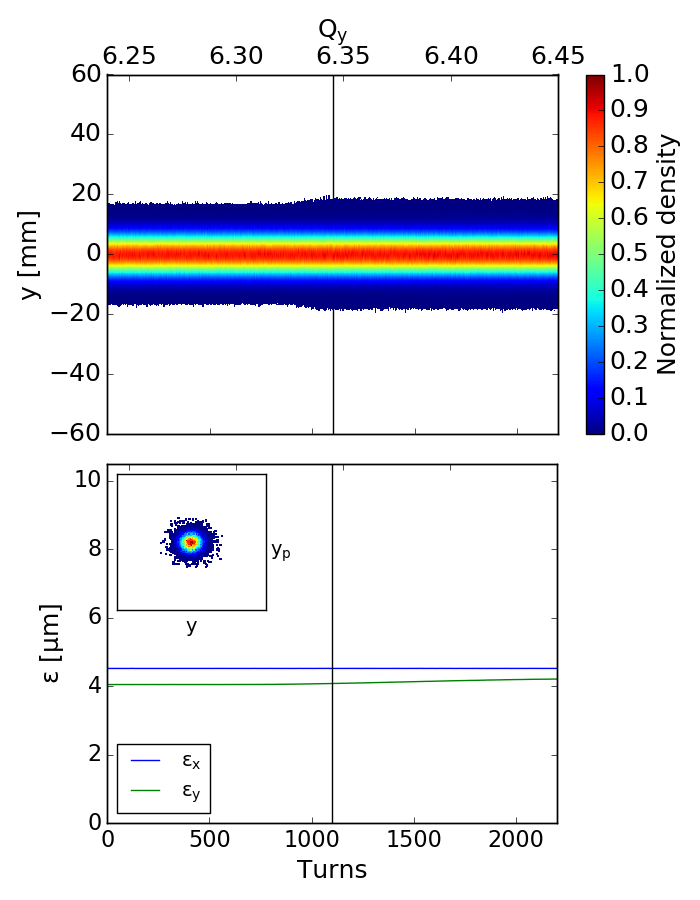}
        \includegraphics[width=0.45\textwidth]{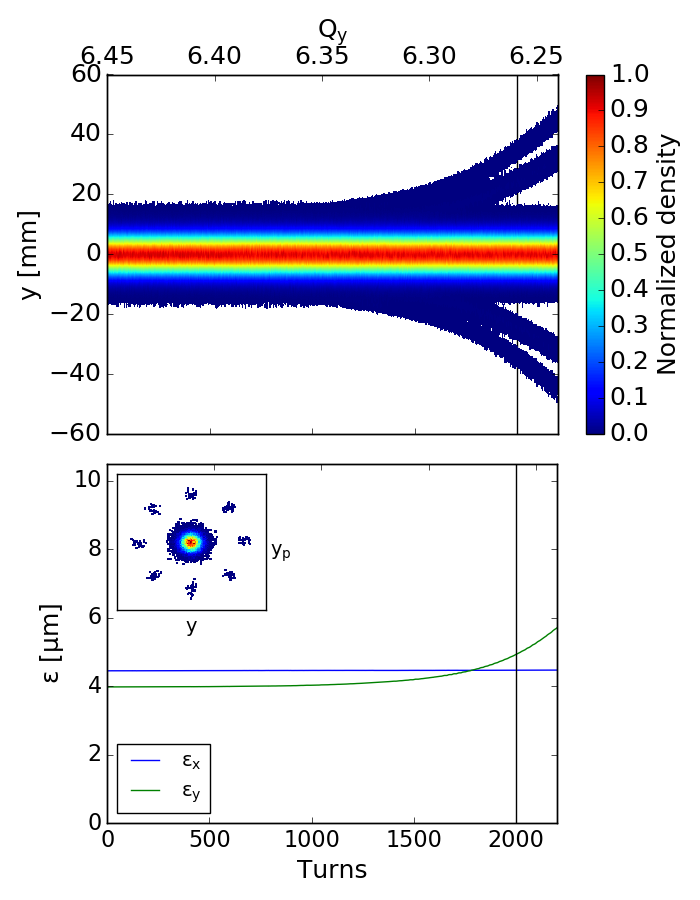}
        \includegraphics[width=0.45\textwidth]{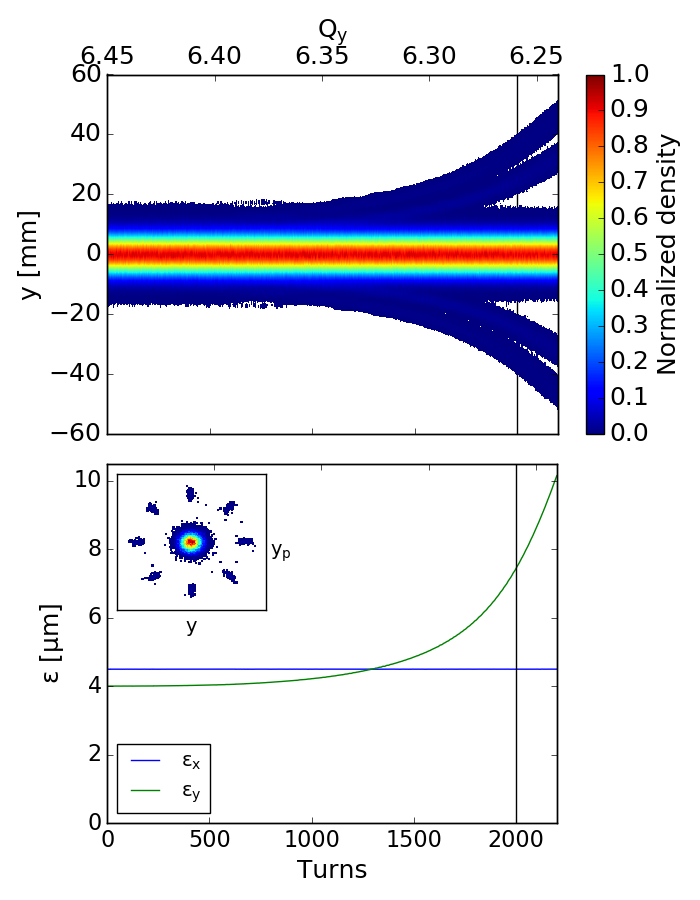}
    \caption{Dynamic crossing of the $Q_y=6.25$ resonance using the simple FODO lattice. The tracking is done using the self-consistent PIC 2.5\,D space charge solver (left) and the frozen model (right). The resonance is crossed as the tune increases from $Q_y=6.24$ to $Q_y=6.45$ (top) and vice-versa (bottom). For each solver and crossing direction, the vertical beam profiles are shown as function of turns and tune, using a second axis, color coded to the particle density (top). Likewise, the transverse emittances are given as a function of turns (and tunes on the second axis), while the vertical phase space is shown color-coded to the particle density for the turn and tune corresponding to the vertical line, at 900 and 2000 turns for the two crossings (bottom).}
    \label{fig:fodoDynamic}
\end{figure*}

Figure~\ref{fig:fodoDynamic} summarizes the simulation results. Hardly any response of the beam is observed when crossing the resonance from below (top graphs) for both the self consistent PIC simulation (left) and the frozen model (right). 
The core of the beam is unaffected while only some minor tails in the vertical plane are formed, probably caused by the scattering of individual particles on the incoherent resonance, which result in a vertical emittance growth of a few percent.
In fact, the phase space does not show any structure that would indicate the crossing of a strong resonance. 

The situation is quite different when crossing the resonance from above (bottom graphs). A large increase of the vertical emittance is observed and the phase space shows the clear formation of 8 islands indicating the excitation of the resonance in $\mathrm{8^{th}}$ order. This explains the formation of large tails, since particles trapped in the $8Q_y=50$ resonance separate from the beam core as the resonance islands move outwards when the vertical tune is further decreased. It is worth pointing out that the qualitative beam behaviour between the fully self-consistent PIC simulation (left) and the frozen space charge solver (right) agrees very well. 
In fact, the main dynamics of the two models is equivalent, which confirms the incoherent nature of the resonance since the frozen space charge solver cannot reproduce coherent effects.
However, the emittance blow-up in the frozen model is slightly more pronounced compared to the self consistent solver. This quantitative difference is due to the fact that the beam parameters used for the frozen potential are kept constant throughout the simulation while the change of the particle distribution is taken into account in the self-consistent PIC simulations. 

\subsection{The PS lattice (coasting beam)}

\begin{figure*}[p] 
    \centering
        \includegraphics[width=0.45\textwidth]{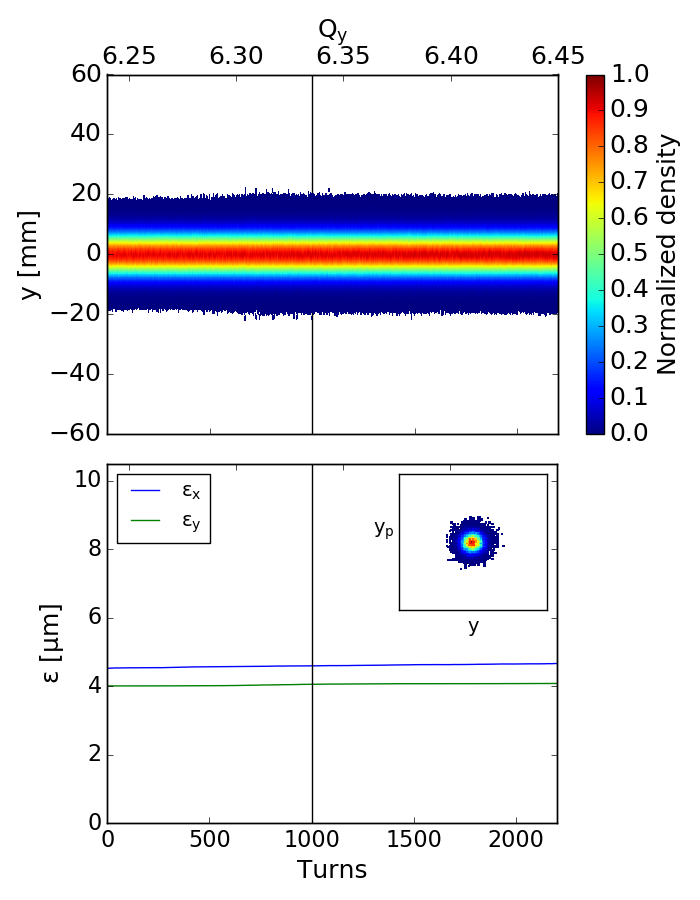}
        \includegraphics[width=0.45\textwidth]{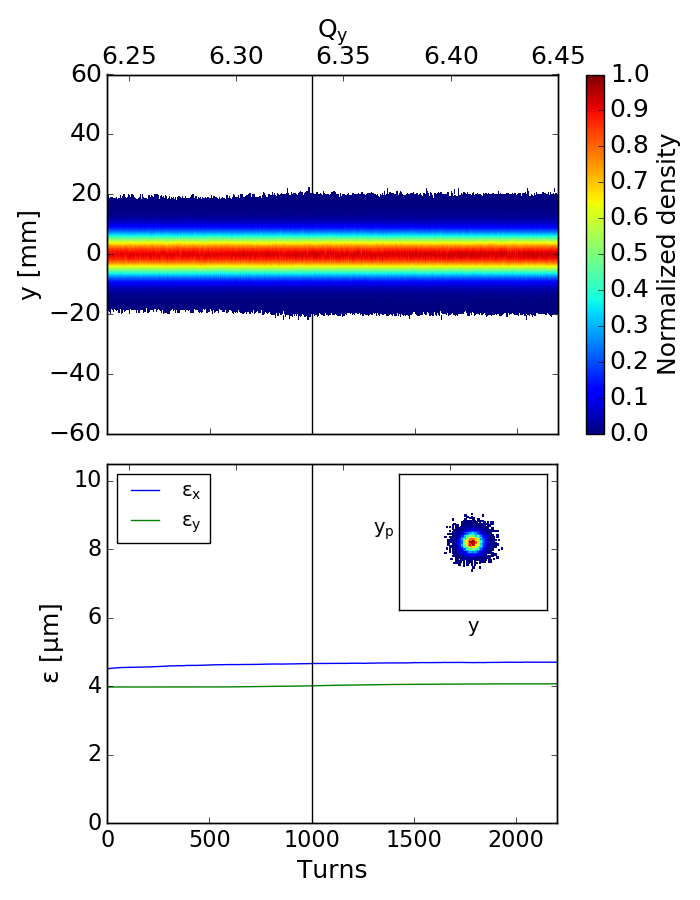}\\
        \includegraphics[width=0.45\textwidth]{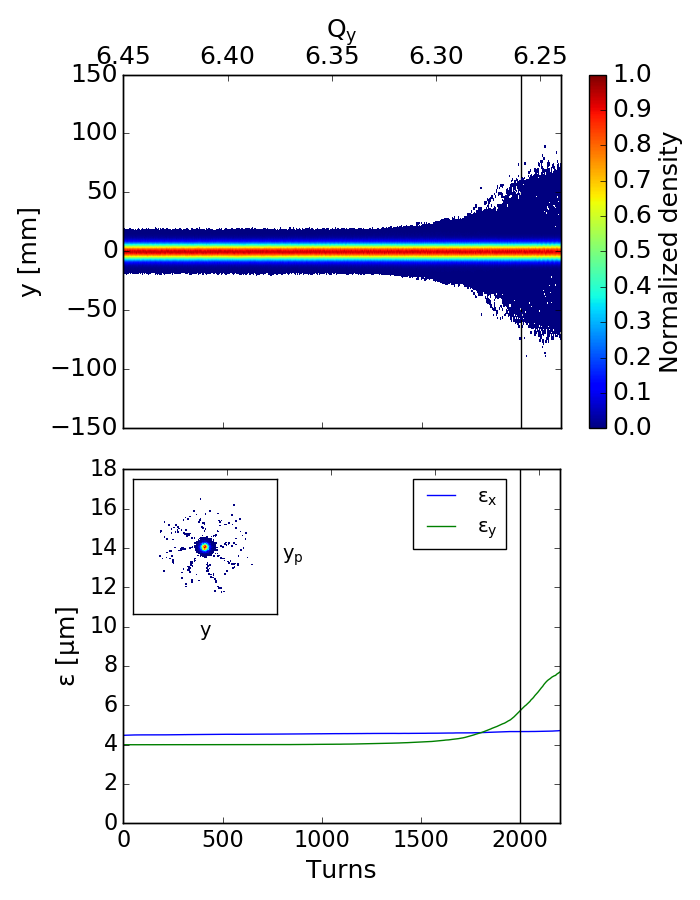}
        \includegraphics[width=0.45\textwidth]{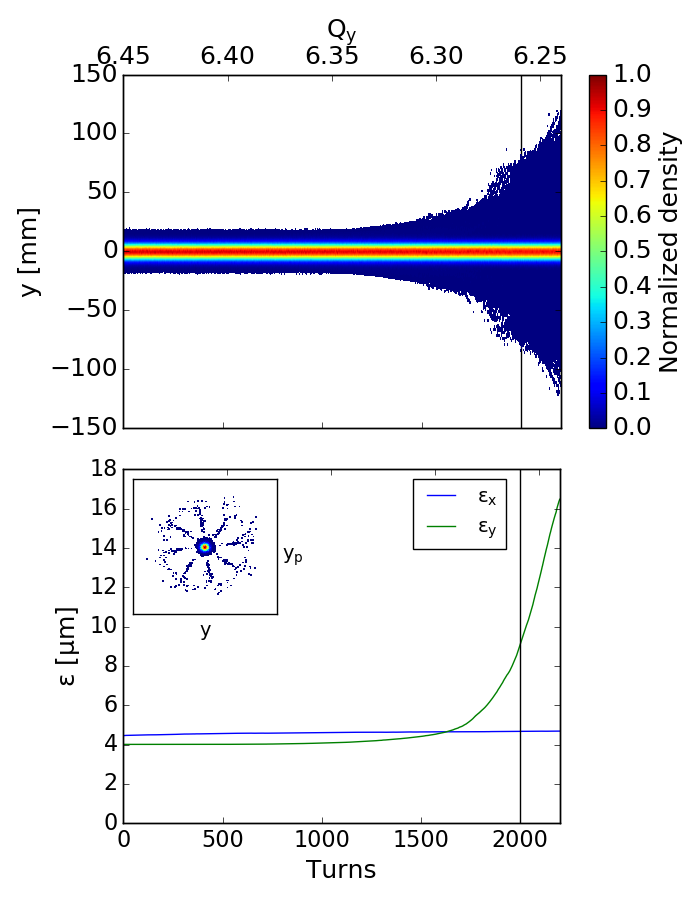}
    \caption{Dynamic crossing of the $Q_y=6.25$ resonance using the PS lattice. The tracking is done using the self-consistent PIC 2.5 \,D space charge solver (left) and the frozen model (right). The resonance is crossed as the tune increases from $Q_y=6.24$ to $Q_y=6.45$ (top) and vice-versa (bottom). For each solver and crossing direction, the vertical beam profiles are shown as function of turns and tune, using a second axis, color coded to the particle density (top). Likewise, the transverse emittances are given as a function of turns (and tunes on the second axis), while the vertical phase space is shown color-coded to the particle density for the turn and tune corresponding to the vertical line, at 900 and 2000 turns for the two crossings (bottom). Note the different scale of both profiles and emittance plots between the top and bottom graphs.}
    \label{fig:psDynamic}
\end{figure*}

A similar study as shown above for the FODO lattice was repeated using the actual lattice of the PS. 
The main difference between the simple FODO lattice and the PS is that the latter consists of combined function magnets, and therefore the modulation of the lattice functions along the machine is less pronounced. 
In addition, the symmetry of the PS lattice is slightly modified by the 20 long straight sections as described in Sec~\ref{sec:resonance-identification}. 
Although extra quadrupole magnets are installed in the PS, in these simulations the tune matching was done using only the quadrupolar components of the PFW to preserve the periodicity as much as possible. 
The nonlinear field components introduced by the PFW in the real machine were omitted, in order not to excite any additional resonances.
Hence, only the linear PS model is considered with space charge as the only nonlinearity in the lattice.
The results for both crossing directions and the two solvers are shown in Fig.~\ref{fig:psDynamic}.

Similar to the observations for the simple FODO lattice, the response of the beam as the tune increases from $Q_y=6.24$ to $Q_y=6.45$ is minimal and no significant vertical emittance increase is observed (top graphs). 
On the contrary, crossing the resonance in the opposite direction results in significant vertical emittance growth (bottom graphs). 
With both space charge models the vertical phase space shows particles in the tails of the beam distribution trapped in 8 islands that detach from the core, spread out and eventually collapse at high amplitudes.
The profile evolution shows this trapping and collapse as the tails expand to very large amplitudes. 
The fact that both solvers demonstrate the same behaviour confirms again the incoherent nature of this resonance.  

\subsection{Long term behaviour (bunched beam, simple FODO lattice )}

The simulations of the dynamic crossing have shown that in both lattices the incoherent response of the beam is by far dominating, even in the absence of longitudinal motion.
In order to study the long term behaviour in the presence of synchrotron motion, a bunched beam was tracked in the FODO lattice for $\mathrm{10^5}$ turns (corresponding to about \SI{0.2}{\second} and more than 140 synchrotron oscillations in the PS). 
The FODO lattice is chosen, since the PS lattice has significantly more elements and tracking with the PIC solver is unfeasible for so many turns due to the excessive simulation time required.
The beam behaviour at tunes in the vicinity of the working point regime of interest is explored, namely $Q_x=6.2$ and $Q_y=6.24-6.31$.
The scan is static, i.e.~the tunes are kept constant throughout the simulations and the initial beam distributions are Gaussian in the transverse planes while the longitudinal profile was parabolic, similar to the operational beams in the PS.
For the simulations with the PIC solver, $4\times10^{5}$ macro-particles were used, while in the simulations with the analytic solver already $3\times10^{3}$ macro-particles were enough to resolve the beam profiles.
To ensure that the analytic space charge solver is not overestimating the resonance excitation when the beam is degrading, the beam parameters used for calculating the space charge kick (intensity, transverse emittances, longitudinal line density profile and momentum spread) were updated every 100 turns.

\begin{figure}[!t]
    \centering
    \includegraphics[width=0.98\columnwidth]{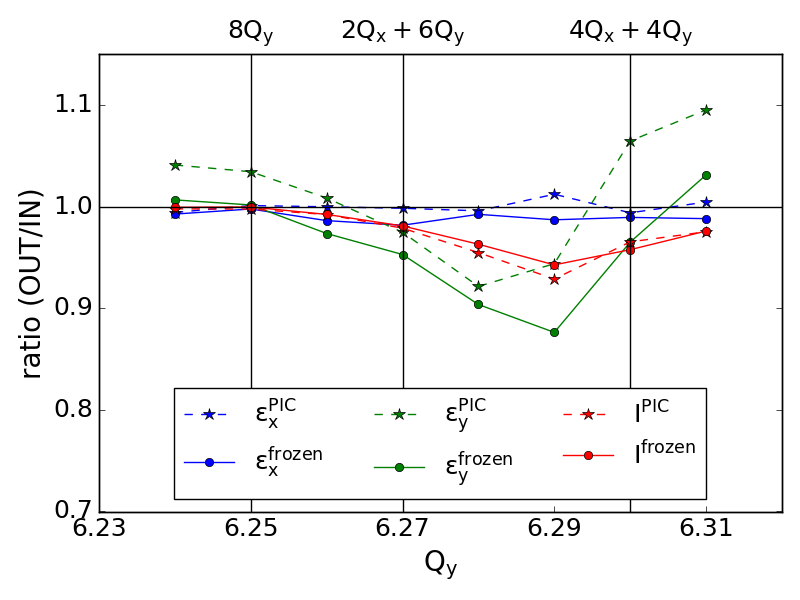}
    \caption{Static tune scan of the simple FODO lattice for $\mathrm{10^5}$ turns of a bunched beam with space charge. Transverse emittances and intensity (I) ratios are plotted as a function of the vertical tune for the self-consistent PIC 2.5\,D and the adaptive frozen space charge solvers with dashed and solid lines, respectively. The vertical lines correspond to $\mathrm{8^{th}}$ order systematic resonances crossed.}
    \label{fig:picFrozen}
\end{figure}

A comparison of the transverse emittance and intensity ratios (ratio final over initial) between the two solvers is shown  in Fig.~\ref{fig:picFrozen}.
The intensity and horizontal emittance values seem to agree very well between the two models for all tested working points.
However, a disagreement of a few percent (up to $\approx9\%$) is observed in the vertical emittances, especially for $Q_y>6.29$.
The larger blow-up observed in the PIC simulations could be partially coming from the noise on the grid~\cite{PhysRevSTAB.17.124201}, and partially from the fact that the Bassetti-Erskine formula used in the analytic solver is describing exact Gaussian distributions.
Overall, the agreement between the solvers is very good.
The analytic solver with parameters updated every 100 turns, which is referred to as adaptive frozen model in what follows, is therefore the simulation model of choice as it provides good enough agreement with the PIC model and allows tracking of less particles and consequently leads to much shorter simulation times.

It should be noted that, besides the agreement between the solvers, the results of this simulation study hint at the excitation of additional space charge driven resonances, indicated by the vertical lines in the graph, which will be discussed in more detail in the following section. 

\section{Resonance Driving Terms}
\label{sec:RDTs}

\begin{figure*}[!tb]
    \centering
	\includegraphics[width=0.48\textwidth]{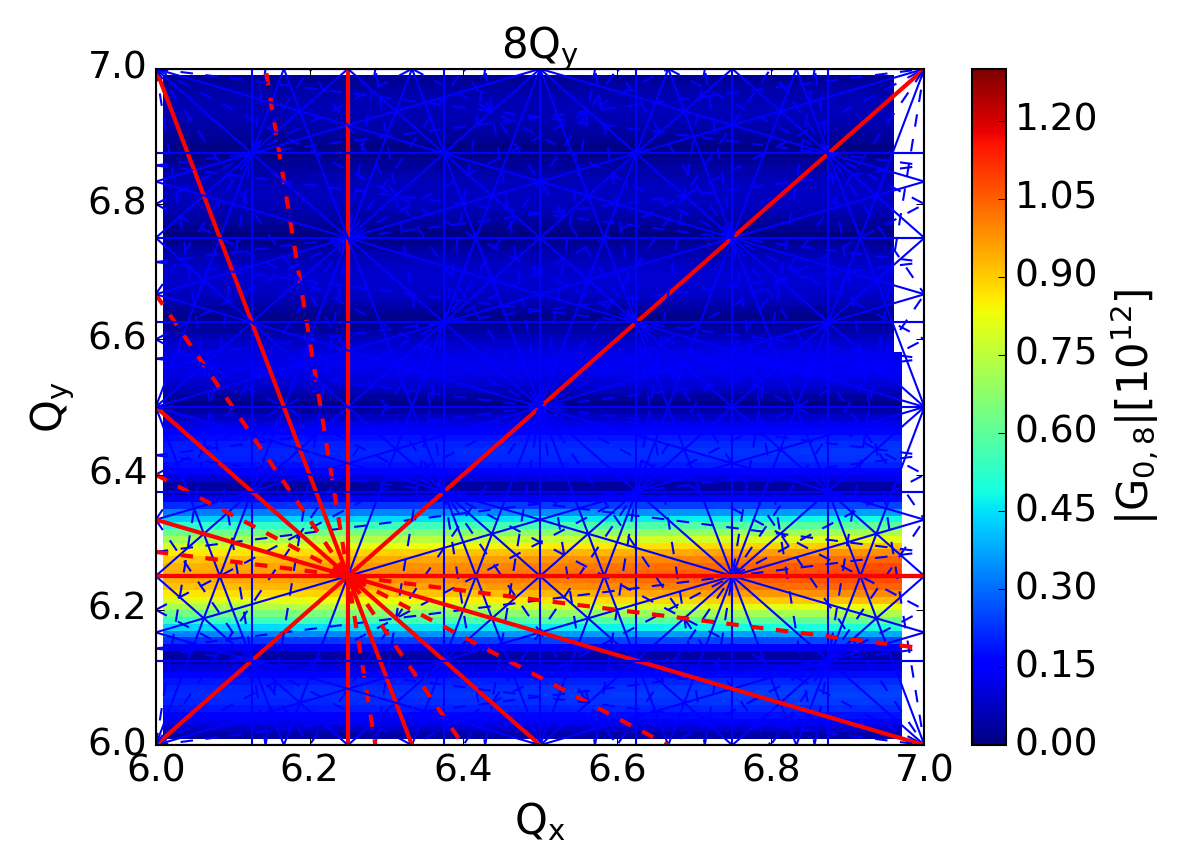}
	\includegraphics[width=0.48\textwidth]{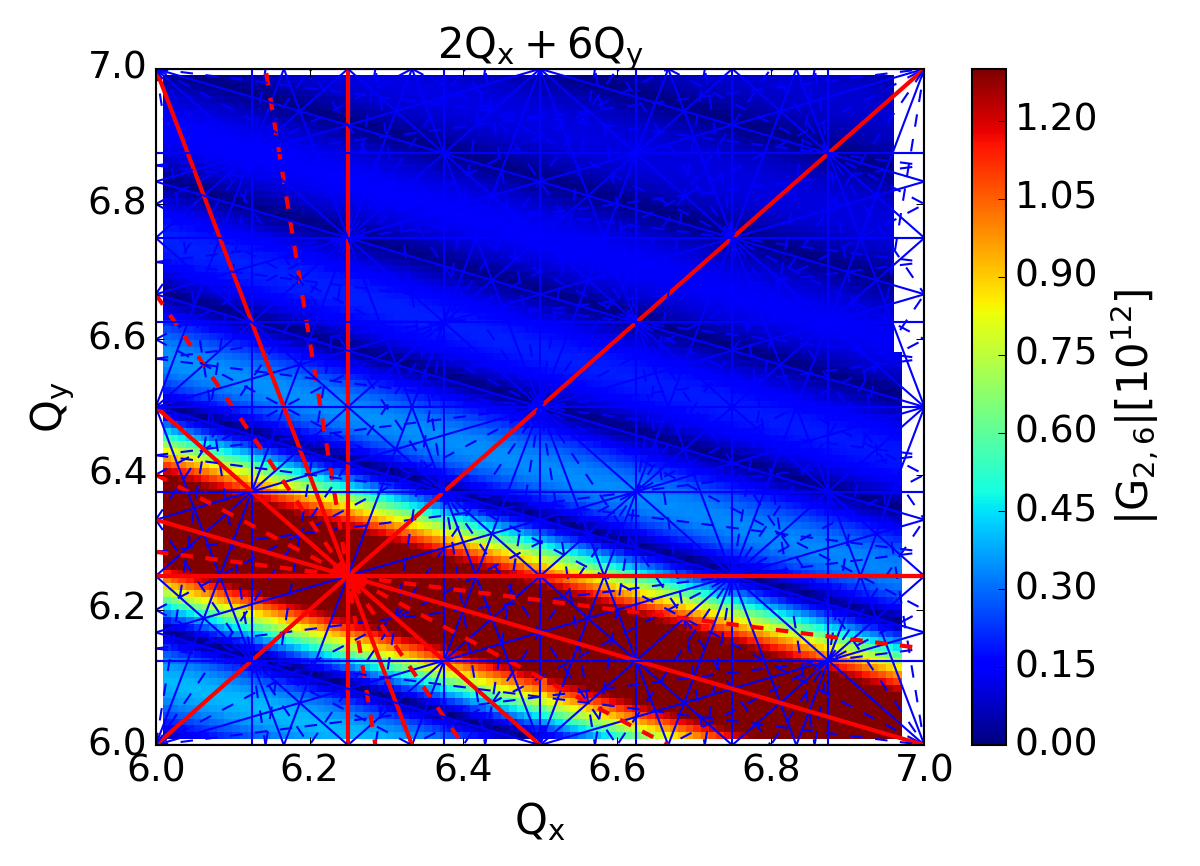}\\
	\includegraphics[width=0.48\textwidth]{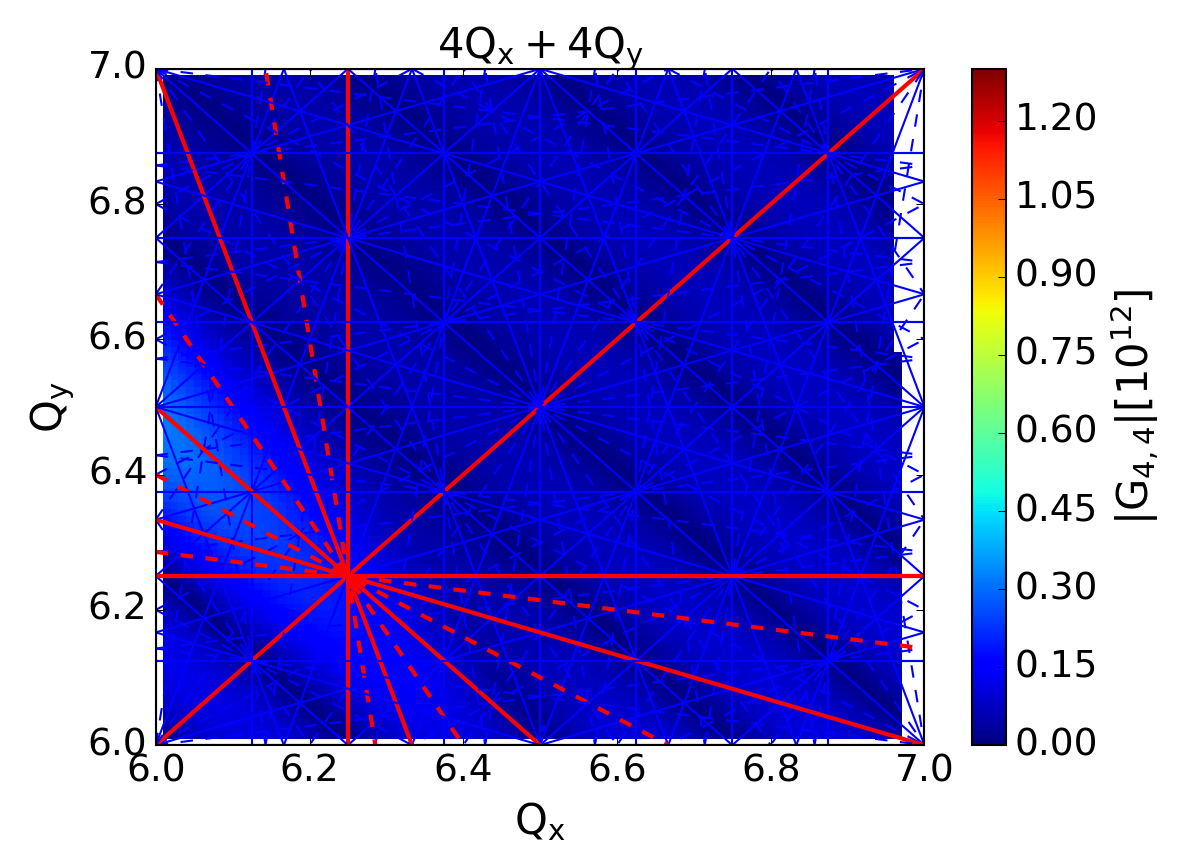}
	\includegraphics[width=0.48\textwidth]{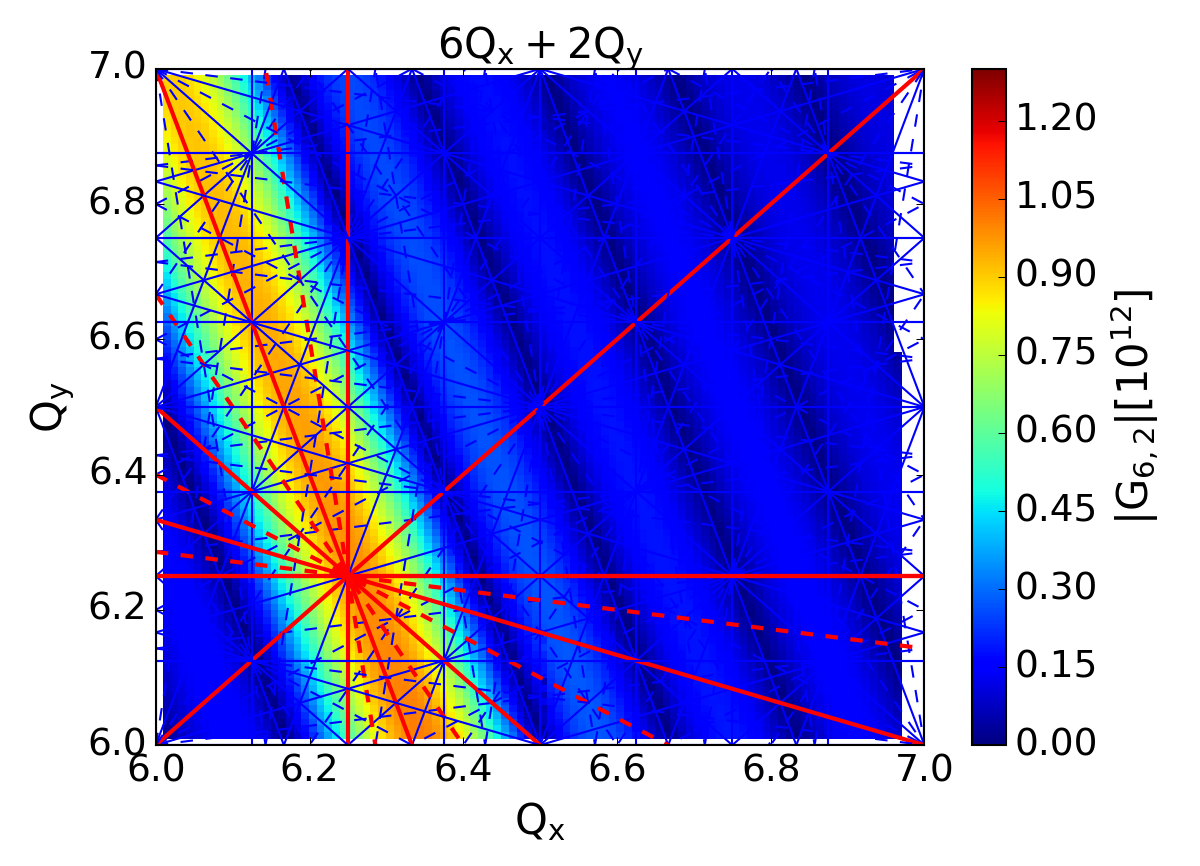}\\
	\includegraphics[width=0.48\textwidth]{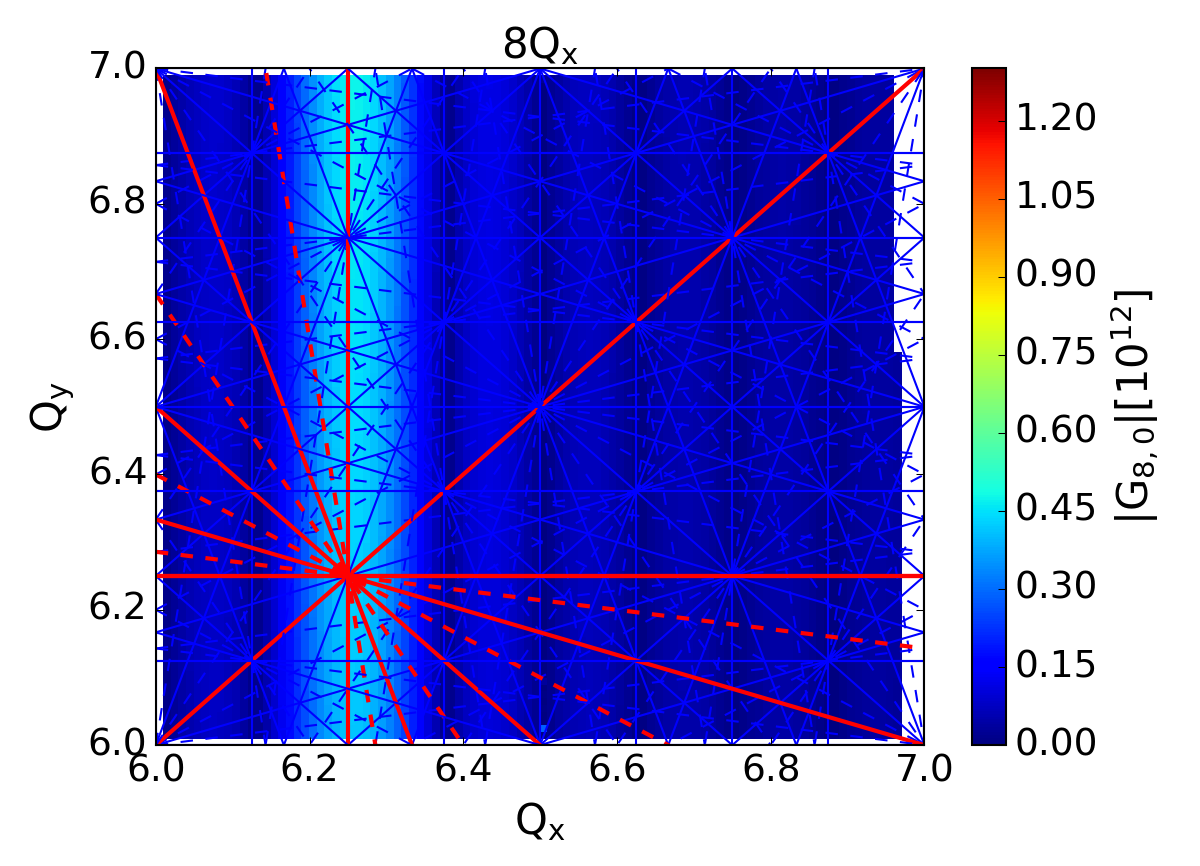}
	\includegraphics[width=0.48\textwidth]{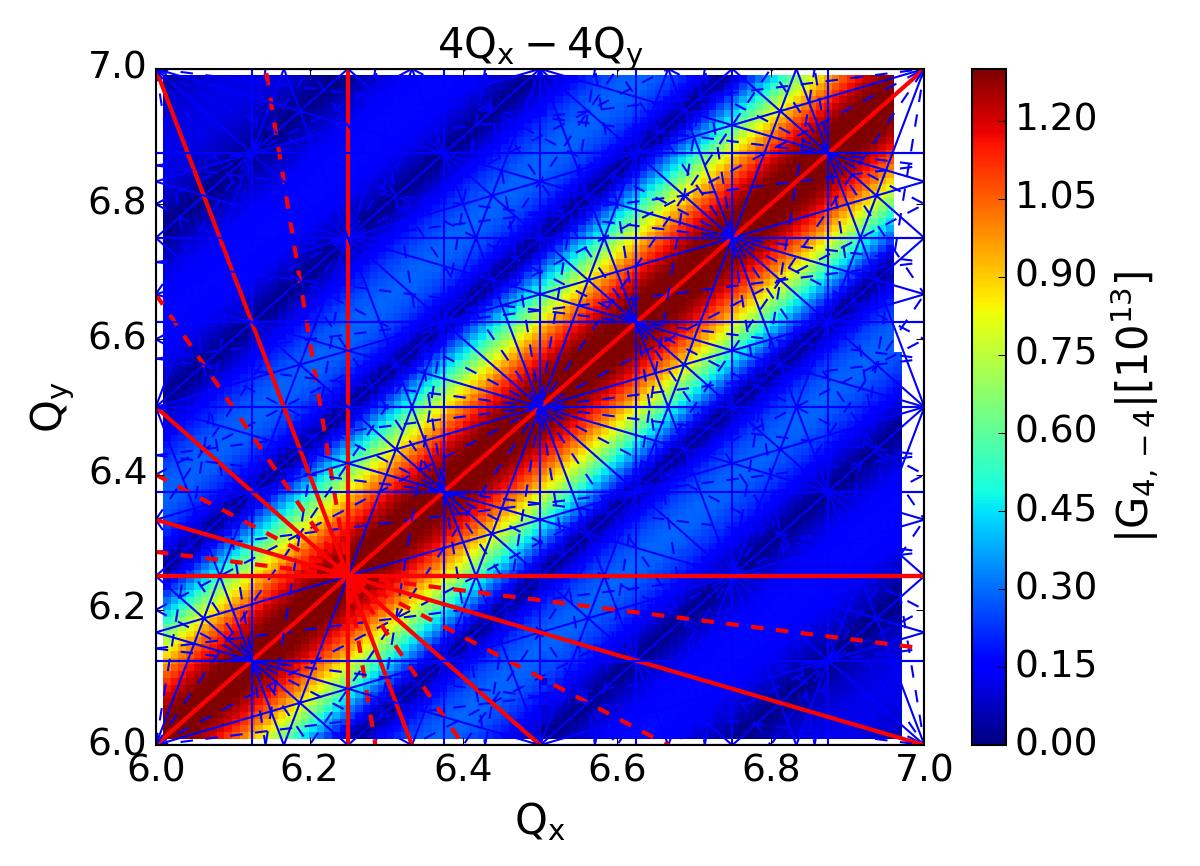}
    \caption{RDTs computed from the space charge potential of a transverse Gaussian beam for the PS lattice for the $8Q_y$ (top-left), $2Q_x+6Q_y$ (top-right), $4Q_x+4Q_y$ (middle-left), $6Q_x+2Q_y$ (middle-right), $8Q_x$ (bottom-left) and $4Q_x-4Q_y$ (bottom-right) resonances for various working points. The color code corresponds to the amplitude of the respective driving term.Resonance lines of $\mathrm{8}^{\mathrm{th}}$ order are plotted. Skew in dashed and normal in solid lines, systematic in red and non-systematic in blue.}
    \label{fig:RDTs}
\end{figure*}

Since the nature of the resonance at $Q_y = 6.25$ is purely incoherent, as revealed by the simulations discussed above, its strength can be studied by applying classical perturbation theory on the nonlinear Hamiltonian of a stationary beam~\cite{Guignard:695942}. In fact, the leading order Resonance Driving Terms (RDTs) of incoherent space charge driven resonances can be calculated~\cite{MACHIDA1997316, Lee_2006} for the perturbation coming from the space charge potential $V_{sc}$ of a Gaussian (bunched) beam, which is given by~\cite{Kheifets} 
\begin{equation}\label{eq:SCpotential}
    V_{sc}(x,y)=\frac{r_0N_b}{\beta^2\gamma^3\sqrt{2\pi}\sigma_s}\int_{0}^{\infty}\frac{-1+\exp{-\frac{x^2}{2\sigma^2_x+t}\frac{y^2}{2\sigma^2_y+t}}}{\sqrt{(2\sigma^2_x+t)(2\sigma^2_y+t)}}dt,
\end{equation}
where $r_0$ is the classical particle radius, $N_b$ the bunch intensity, $\mathrm{\beta, \gamma}$, the relativistic factors and $\sigma_{s,x,y}$ the longitudinal, horizontal and vertical beam sizes, respectively.
The evaluation of the above method was implemented in a Python module~\cite{Asvesta:2696190} to calculate the RDTs and the nonlinear detuning terms for the potential of Eq.~\eqref{eq:SCpotential}.
This code was used to study the $\mathrm{8^{th}}$ order structure resonances in the PS in the vicinity of the operational working point, as discussed below. The optics functions of the PS lattice, required for the calculation of the RDTs, were obtained from MAD-X.
A detailed explanation of the driving term calculation can be found in Appendix~\ref{app:RDTs}.

The RDTs for the $\mathrm{8^{th}}$ order resonances in the full tune space around the nominal working point of the PS, namely $8Q_y=50$, $2Q_x+6Q_y=50$, $4Q_x+4Q_y=50$, $6Q_x+2Q_y=50$, $8Q_x=50$ and $4Q_x-4Q_y=0$,
were evaluated for a high brightness beam with the parameters given in the first column of Table~\ref{tab:parameters} and the results are shown in Fig.~\ref{fig:RDTs}. The coupling resonance $4Q_x-4Q_y=0$ is driven in $\mathrm{8^{th}}$ order through the $\mathrm{0^{th}}$ harmonic and is the strongest resonance of the ones discussed here. Note that this resonance is also excited in $\mathrm{4^{th}}$ order, (i.e.~Montague resonance~\cite{Montague}), but for the purposes of this study only the $\mathrm{8^{th}}$ order resonance driving term is shown.
The $\mathrm{8^{th}}$ order driving terms for the remaining resonances,
 increase significantly when the tune values are set on the $\mathrm{50^{th}}$ harmonic.
This harmonic coincides with the periodicity of the PS optics modulation and consequently, the corresponding resonances are excited.
On the other hand, the resonances at harmonics not present in the variation of the optics functions of the lattice are not excited and the respective RDT goes to 0.
It should be emphasized that the strength of the excitation, indicated by the amplitude of the RDT, is not the same for all resonances.
The $8Q_y=50$, $2Q_x+6Q_y=50$ and $6Q_x+2Q_y=50$ resonances appear to be the strongest ones,
while the $8Q_x=50$ and especially the $4Q_x+4Q_y=50$ resonances are much weaker. The difference comes from the different beam sizes in the horizontal and vertical planes due to the emittances and the presence of dispersion, and from the different parts of the space charge potential driving them.

\begin {table}[!t]
\begin{center}
\caption{Beam parameters used in tune scans, the evaluation of the space charge RDTs and the FMAs.}
    \centering
    \begin{tabular}{ L{3.3cm} C{2.4cm} C{2.4cm}}
    \toprule
     Beam Type& high-brightness & medium-brightness \\ 
    \colrule
    Intensity [\SI{e10}{ppb}] & \SI{96}{} & \SI{48}{} \\
    $ E_\textrm{kin} [\SI{}{GeV}]$ &\SI{1.4}{}& \SI{1.4}{} \\ 
    Bunch length ($4\sigma$)~[\SI{}{ns}] & 80 & 80\\
    $\Delta p \slash p~(\textrm{rms})~[\SI{e-3}{}]$ &\SI{0.52}{}& \SI{0.52}{} \\     
    $\varepsilon_{\mathrm{x}}^{\mathrm{n}}~(1\sigma)$ [\SI{}{\micro\meter}] & \SI{5.5}{} & \SI{5.5}{} \\
    $\varepsilon_{\mathrm{y}}^{\mathrm{n}}~(1\sigma)$ [\SI{}{\micro\meter}]  & \SI{4.5}{} & \SI{4.5}{} \\ 
    $\Delta Q_x$ (maximum) & -0.15 & -0.07  \\ 
    $\Delta Q_y$ (maximum) & -0.16 & -0.08  \\
  \toprule
    \end{tabular}
\label{tab:parameters}

\end{center}
\end{table}

\section{Frequency Map Analysis}
\label{sec:fma}

\begin{figure*}[!t]
    \centering
        \includegraphics[width=0.46\textwidth]{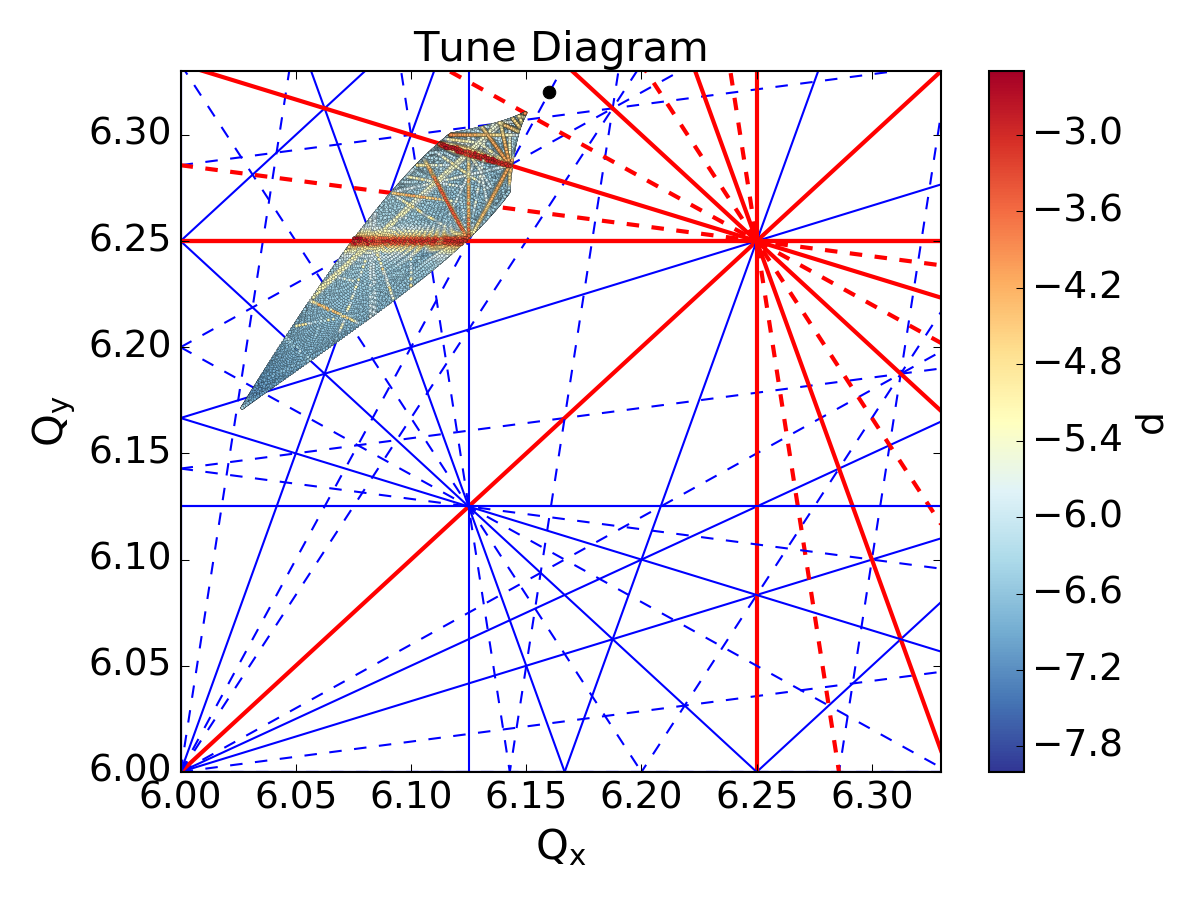}
        \includegraphics[width=0.46\textwidth]{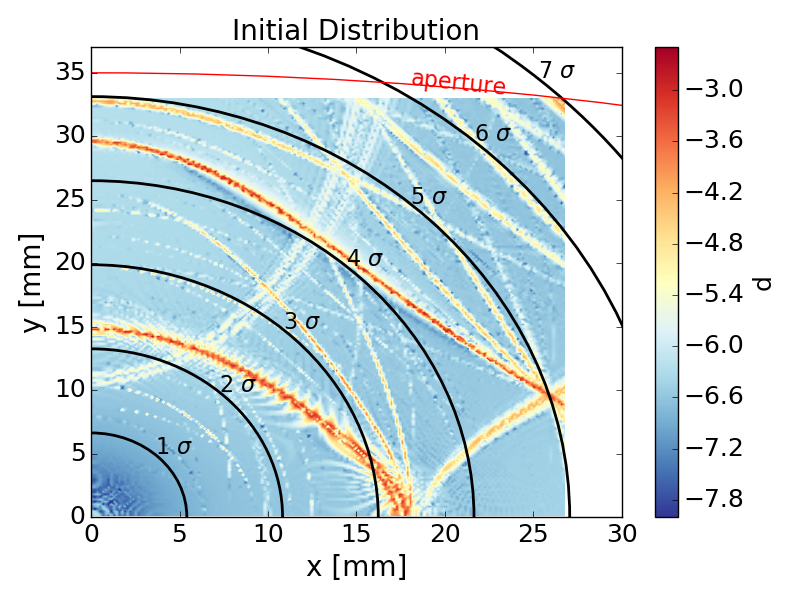}
        \includegraphics[width=0.46\textwidth]{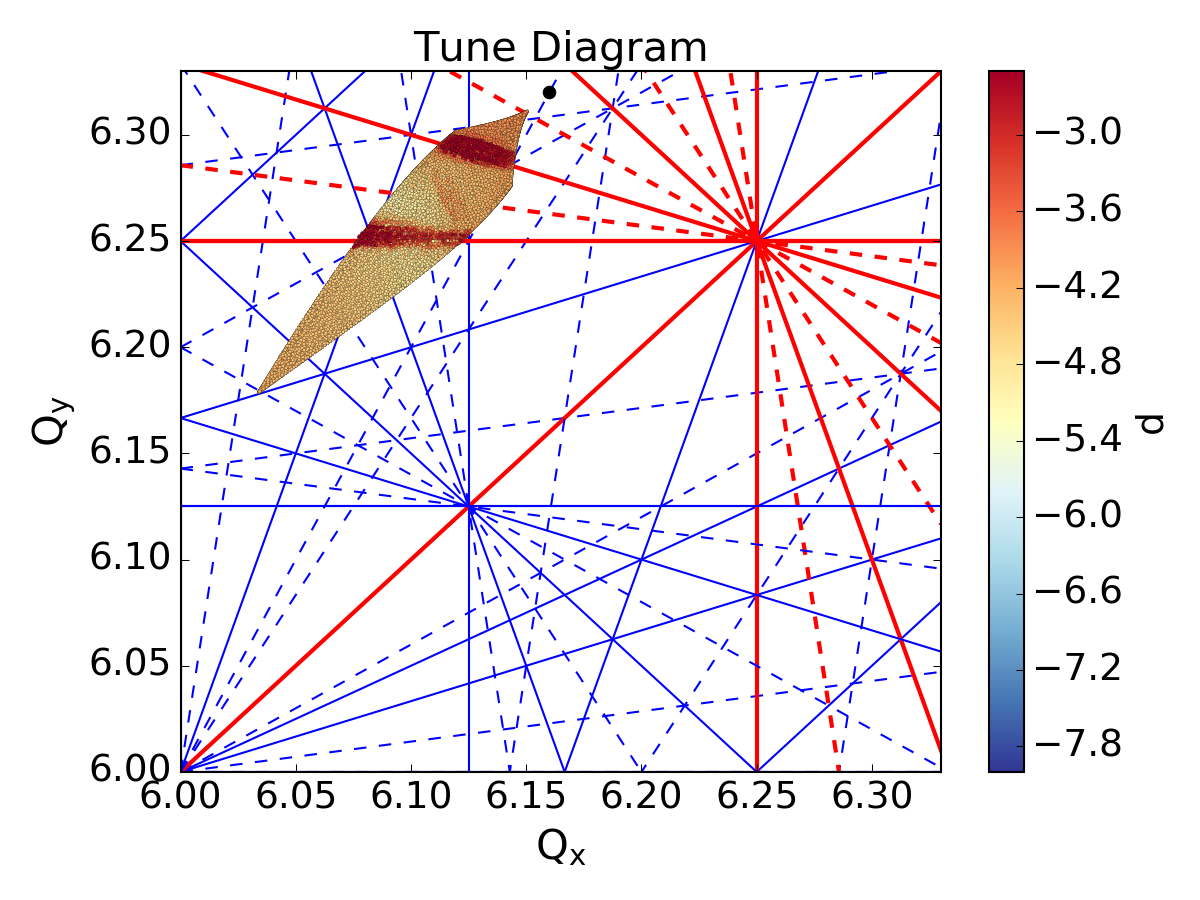}
        \includegraphics[width=0.46\textwidth]{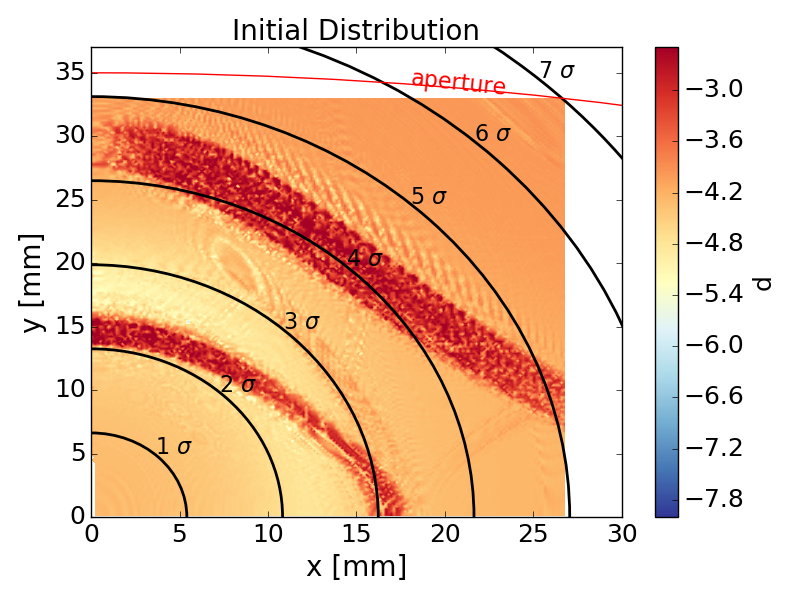}
    \caption{FMA for on-momentum (top) and off-momentum (bottom) particles. Tune diagrams with resonances of $\mathrm{3}^{\mathrm{rd}}$ and $\mathrm{8}^{\mathrm{th}}$ order. Systematic resonances are plotted in red and non-systematic ones  in  blue. The  skew  resonances  are  shown  with  dashed lines and the normal resonances with solid lines (left). The initial position of the particles tracked in the configuration space (right). The color coding represents the diffusion coefficient.}
    \label{fig:fma}
\end{figure*}

The excitation of the space charge driven resonances can also be studied using the Frequency Map Analysis (FMA) technique~\cite{YannisChaos} for tracking data including space charge.
The PS lattice is matched using MAD-X and the tracking is done with PTC in PyORBIT.
The space charge is included using the PyORBIT frozen model without update.
The parameters of the high brightness beam in the first column of Table~\ref{tab:parameters} are used for the calculation of the space charge kicks in the simulations.
For each FMA, test particles with the same longitudinal action are tracked for two consecutive synchrotron periods.
The tunes of each particle are calculated applying the Python implementation of the Numerical Analysis of Fundamental Frequencies (NAFF)~\cite{LASKAR1992253, PyNAFF} to the turn by turn data.
The indicator for the resonance excitation is the tune diffusion coefficient~\cite{LASKAR}, which is defined as $d=\log_{10}\sqrt{{\left(Q_{x,2}-Q_{x,1}\right)}^2+{\left(Q_{y,2}-Q_{y,1}\right)}^2}$, where the indices $(1),(2)$ refer to the first and second synchrotron period, respectively.
For on-momentum particles, the resulting FMA is shown in Fig.~\ref{fig:fma} (top).
The excitation of the structural resonances is shown in the tune and configuration spaces.
The same technique is applied to off-momentum particles, initialized longitudinally at $\approx1.2\sigma$, and the resulting FMA is shown in Fig.~\ref{fig:fma} (bottom).
In this case the tunes are modulated through the synchrotron motion due to the varying space charge potential along the longitudinal line density profile of the beam and the chromaticity, which is kept at the natural values.
Therefore, the resonances appear broader compared to the on-momentum case, since they are crossed by a larger number of particles through synchrotron motion.
Additionally, the calculated tune diffusion for all particles crossing resonances appears larger.

\section{Loss Mechanism}
\label{sec:lossmechanism}

The excitation of the space charge driven resonances due to the $\mathrm{50^{th}}$ harmonic of the lattice functions (cf.~Fig.~\ref{fig:psbeamSize}) has been confirmed in the simulations shown in Fig.~\ref{fig:psDynamic}, the analytical calculation of the driving terms shown in Fig.~\ref{fig:RDTs} and the FMAs of Fig.~\ref{fig:fma} as discussed above.
To identify the underlying mechanism that leads to the beam loss observed in the measurements presented in Figs.~\ref{fig:intensity} and~\ref{fig:tunescanRay}, 
test particles were simulated using the frozen space charge solver without update.
The test particles were initialized with relatively small horizontal initial amplitudes, $x<\SI{10}{\milli\meter}$, and varying vertical initial amplitudes from $y=\SI{6}{\milli\meter}$ to $y=\SI{20}{\milli\meter}$, i.e.~up to $5\,\sigma$ of the beam size.
Longitudinally, all particles were initialized with the same action and in-phase, at $\approx1.2\,\sigma_s^{\mathrm{rms}}$ and $dE=0$. 
The simulations were performed for a working point above the $8Q_y=50$ resonance, namely $Q_x=6.18$, $Q_y=6.28$, and particles were tracked for 2000 turns corresponding to $\approx2.5$ synchrotron periods.

\begin{figure}[!t]
    \centering
        \includegraphics[trim=10 0 10 0, clip, width=0.95\columnwidth]{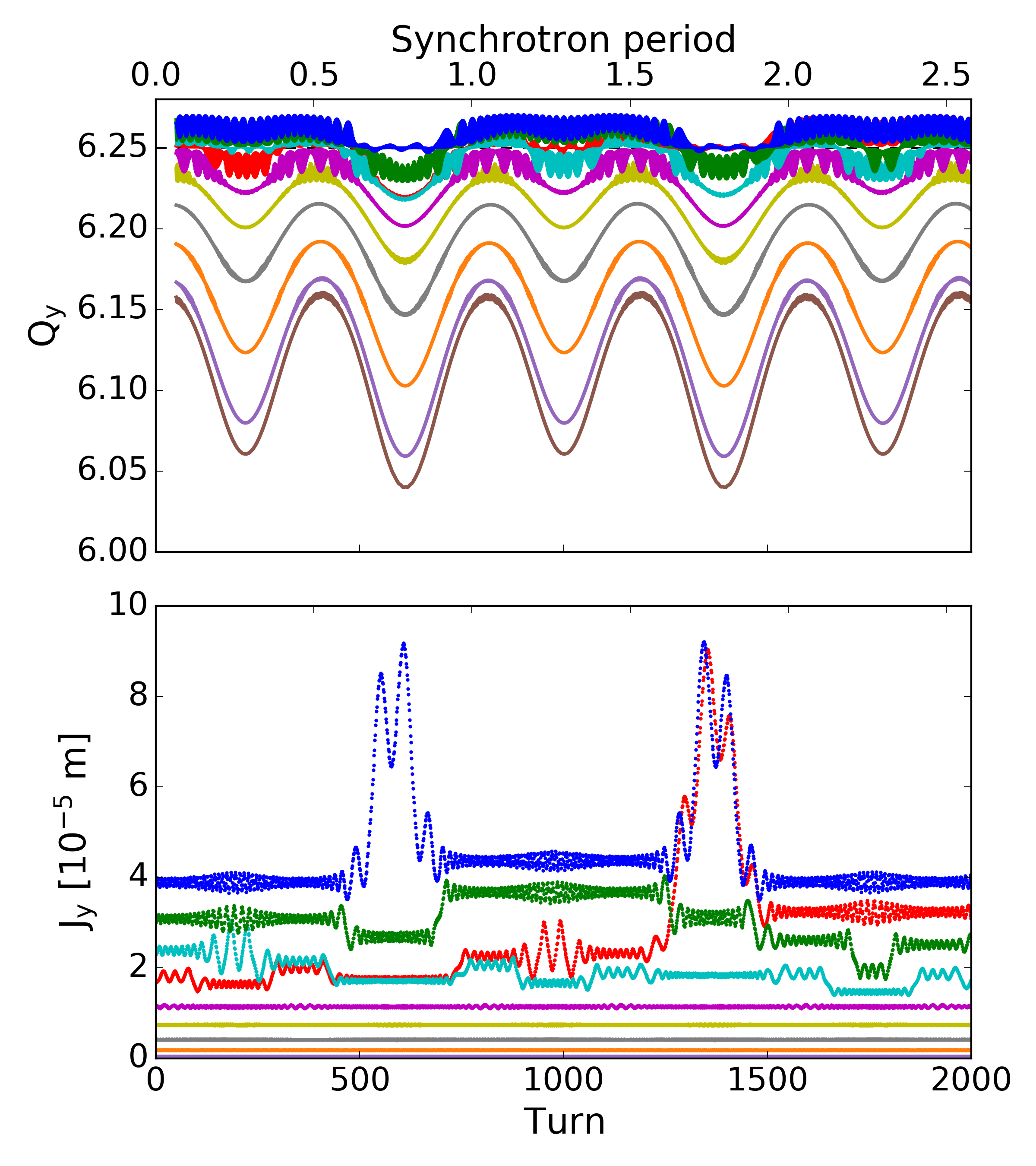}
	\caption{Evolution of the Vertical tune (top) and the vertical betatron action (bottom) of test particles in tracking simulations. The second axis on top indicates the number of synchrotron periods. 
	}
    \label{fig:tunes}
\end{figure}

\begin{figure}[!t]
    \centering
        \includegraphics[trim=10 0 10 0, clip,width=0.85\columnwidth]{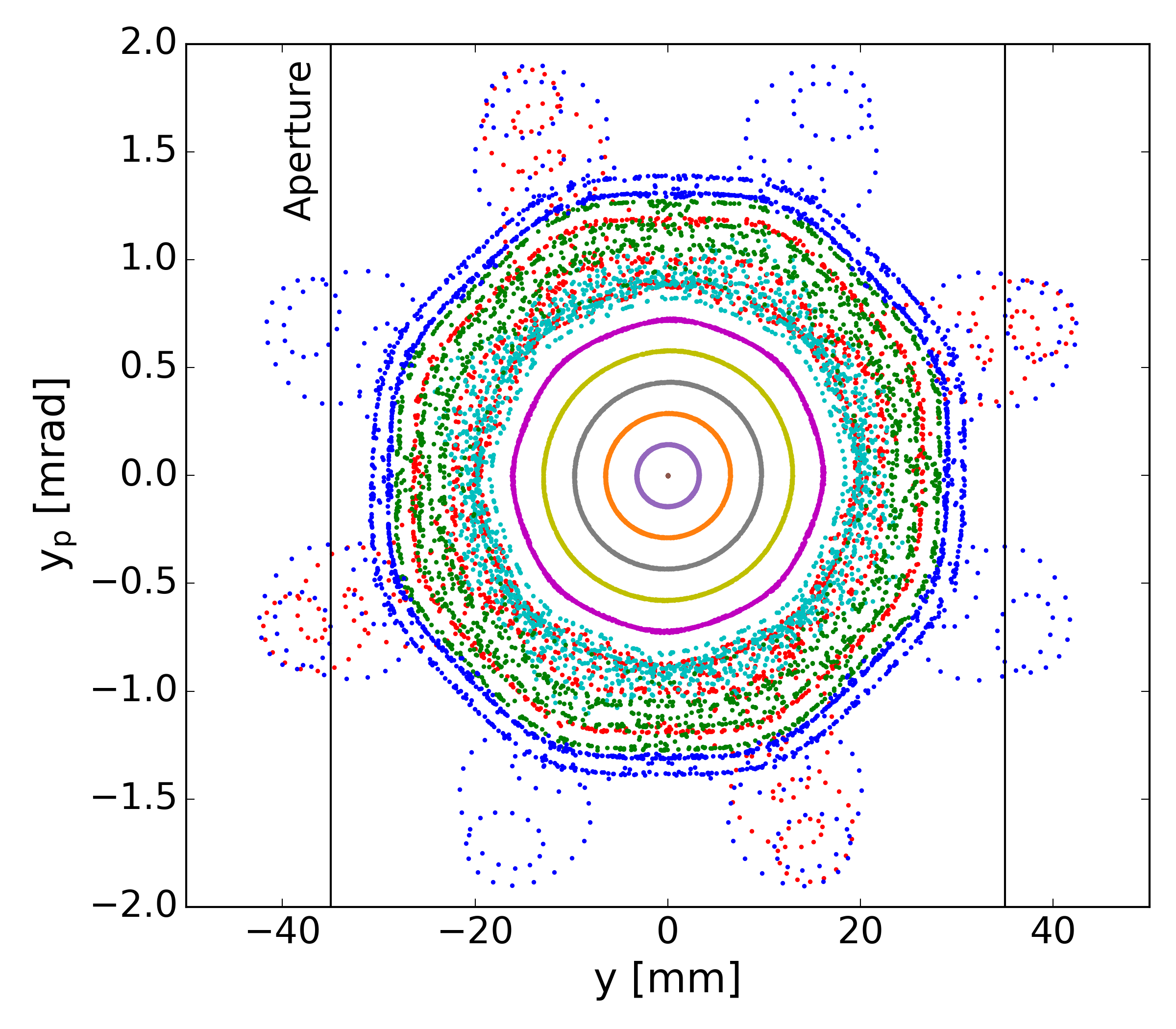}
	\caption{Vertical phase space of the test particles shown with the same color code as in Fig.~\ref{fig:tunes}.}
    \label{fig:verticalPhaseSpace}
\end{figure}
The turn-by-turn data of the simulations were used to calculate the tunes of each particle as described in Section~\ref{sec:fma}.
The tunes were evaluated using a sliding window of 50 turns and their evolution versus the number of turns and synchrotron period is shown in Fig.~\ref{fig:tunes} (top).
The tune modulation is characteristic of the dependence of the transverse space charge force on the longitudinal line density. An additional contribution to the tune modulation comes from chromaticity, which is kept at the natural values of $Q'_{\mathrm{x}}\approx-5.7$ and $Q'_{\mathrm{y}}\approx-7.6$ in the simulations.
Under these conditions the tune modulation is $\Delta Q = \Delta Q_{sc} +\Delta Q_{chroma}$, where the $\Delta Q_{chroma}$ varies from \SIrange{-0.01}{0.01}{} depending on the synchrotron motion.
Hence, in the first half of the synchrotron period the chromaticity gives a positive tune shift while in the second half a negative one, similar to space charge, which results in a different tune modulation depth during the two halfs of each synchrotron period.
Figure~\ref{fig:tunes} (bottom) shows the evolution of the vertical betatron action for the same test particles using the same color code. 
As expected, particles with smaller amplitude have lower tunes due to the incoherent space charge tune shift. 
More interestingly, it turns out that the ``space charge induced periodic resonance crossing'', which was extensively studied in the context of magnet driven resonances in presence of space charge~\cite{ref:Franchetti2003,ref:Metral2006, ref:Franchetti2006, ref:Franchetti2017}, is also the mechanism resulting in losses here: 
for the chosen vertical machine tune, the test particles with high amplitude periodically cross the $8 Q_y=50$ resonance. 
In some cases the crossing of the resonance leads to a rapid diffusion also referred to as "scattering" of the particle trajectory on the resonance~\cite{ref:Franchetti2006}, as can be observed by a change of the particle's action.
In other cases, the resonance crossing leads to a trapping of the particle on the resonance islands so that the particle actions are transported to large amplitudes. 

The trapping can be nicely observed in Fig.~\ref{fig:verticalPhaseSpace}, which shows the vertical phase space portrait for the same test particles using the same color code as before. 
The phase space clearly shows the formation of 8 islands, similar to Fig.~\ref{fig:psDynamic}. 
Trapped particles follow a spiraling trajectory with increasing action in phase space, since the position of the resonance islands is moving to higher amplitudes due to the increasing space charge detuning when the particle is approaching the longitudinal center of the bunch during its synchrotron oscillation. 
Note that the trapped particles leap into every second resonance island due to the fractional tune of $0.25$. 
This simulation was performed without mechanical aperture limitations in order to illustrate the particle dynamics.
However, in the real machine the particles trapped on the resonances islands would be lost when reaching the vertical machine aperture limitation, which is indicated by the vertical lines in the phase space plots. 

The betatron phase of the particle at the moment of the resonance crossing determines in which set of islands a particle is trapped in. Furthermore, beam loss in the machine will occur continuously, as the particles have different longitudinal phases, in agreement with the experimental observations shown in Fig.~\ref{fig:intensity}. 
It should also be emphasized that, in the absence of any positive detuning sources such as chromaticity, the losses due to the $8Q_y=50$ resonance occur only for machine working points with $Q_y>6.25$, since only then the particle tunes are shifted towards the resonance and thus can get trapped and scattered at the resonance islands due to the space charge induced periodic resonance crossing. 

\section{Detailed Machine Studies}
\label{sec:machinestudies}

Detailed machine experiments were performed in order to study the interplay between space charge and the resonances excited either by space charge itself or other lattice nonlinearities.
The measurements consist of static tune scans, during which the beam was injected directly on the matched tunes and stored for $\SI{1.2}{s}$ at injection energy.
The intensity along the cycle was recorded and compared to macro-particle simulations using the adaptive frozen space charge solver.

The tune scan shown in Fig.~\ref{fig:loss_varying_brightness} had already suggested the correlation of the beam brightness to measured losses, while other measurements~\cite{Asvesta:2301799, Asvesta_2018} suggested the presence of multiple $\mathrm{8}^{\mathrm{th}}$ order structure resonances around the operational working point of the PS.
The present study aims to clearly show the presence of these resonances~\cite{Asvesta_2019} and establish the source of excitation by correlating the beam loss and the beam brightness not only to the space charge driven resonances but additional controlled lattice driven ones.

In this respect, two different types of beams were used for the measurements, namely a high-brightness and a medium-brightness beam.
The parameters for both beams are summarized in Table~\ref{tab:parameters}. 
It should be emphasized that 
the longitudinal parameters as well as the transverse emittances and consequently the beam sizes were kept the same for both beams. 
The change of the bunch intensity results in a proportional change of the beam brightness and the space charge force. 
This choice ensures that the RDTs, indicating the resonance strength, are only scaled with the intensity while the relative strength of the excitation between the resonances remains unaffected.
Hence, the different response of the beams on the space charge driven resonances can be directly associated to the change in the space charge force, providing a clear correlation between resonance strength and brightness.
To explore the dependence of the beam losses on the beam brightness on a magnet error driven resonance, the $3Q_y=19$ was excited in a controlled manner.
In fact, a different behaviour of the losses along the controlled resonance compared to the losses from space charge driven resonances can further help to classify the various resonances. 
As the $3Q_y=19$ is naturally excited in the PS as shown in Fig.~\ref{fig:alex}, this resonance had to be compensated before it could be used as a controlled resonance. 
The compensation procedure is described in Appendix~\ref{app:compensation}.

\begin{figure}[!tb]
    \centering
    \includegraphics[clip, trim=2.cm 0cm 3.cm 0cm, width=0.97\linewidth]{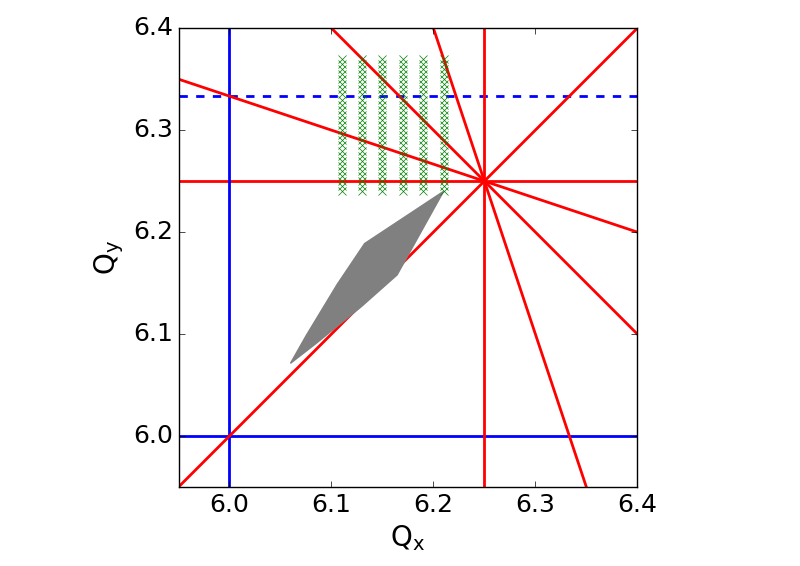}
    \includegraphics[clip, trim=2.cm 0cm 3.cm 0cm, width=0.97\linewidth]{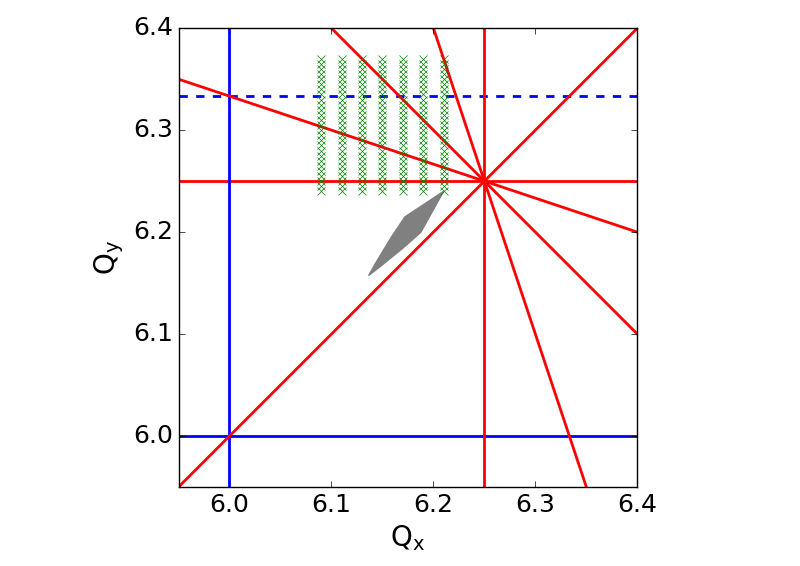}
    \caption{Analytically calculated tune footprints at the nominal working point for the high-brightness (top) and medium-brightness (bottom) beams. The normal $\mathrm{8}^{\mathrm{th}}$ order resonances are shown in red, the $\mathrm{3}^{\mathrm{rd}}$ order skew resonance is indicated by the dashed blue line and the solid blue lines correspond to the integer resonances.
	The green markers correspond to the working points studied in the experiment.}
        \label{fig:tunescan}
\end{figure}

The incoherent space charge tune spreads for both beams were calculated analytically from the potential~\cite{Asvesta:2696190} and are illustrated in Fig.~\ref{fig:tunescan}, together with the tune space covered in the measurements and the resonances of interest.
The tested working points range from 6.24 to 6.37 in steps of $2\times10^{-2}$ in $Q_x$, and 6.11 to 6.21 in steps of $5\times10^{-3}$ in $Q_y$ for the high-brightness beam, while
for the medium-brightness the scan is extended to even lower tune values in $Q_x$, i.e.~6.09 to 6.21.
The reason for this is that the medium-brightness beam with its smaller tune shift could be injected at lower horizontal tunes, while remaining unaffected by resonances at the integer tune of $Q_x=6.0$. 
The beam loss was determined using the intensity values measured with the beam current transformer at \SI{15}{ms} and at \SI{1115}{ms} after injection.
To keep the PS lattice as linear as possible, the chromaticity was kept at the natural values, $Q'_{\mathrm{x}}\approx-5.7$ and $Q'_{\mathrm{y}}\approx-7.6$, the linear coupling was corrected using the closest tune approach method and the transverse damper was used to stabilize the beam~\cite{Sterbini:2158994} from the  head-tail instability on the injection plateau ~\cite{Metral:1051102}.

\subsection{Tune Scans with $\mathbf{3Q_y}$ Resonance Compensated}

\begin{figure*}[!tb]
    \centering
        \includegraphics[width=0.494\textwidth]{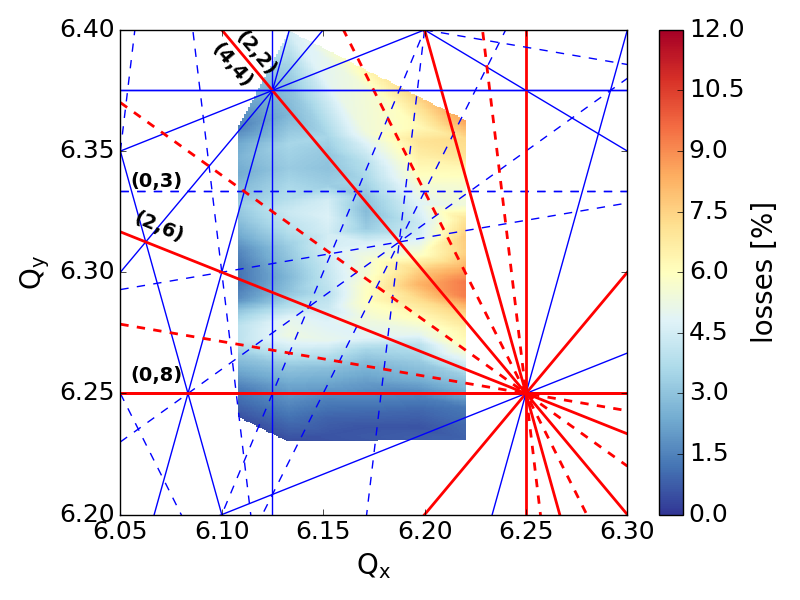}
        \includegraphics[width=0.494\textwidth]{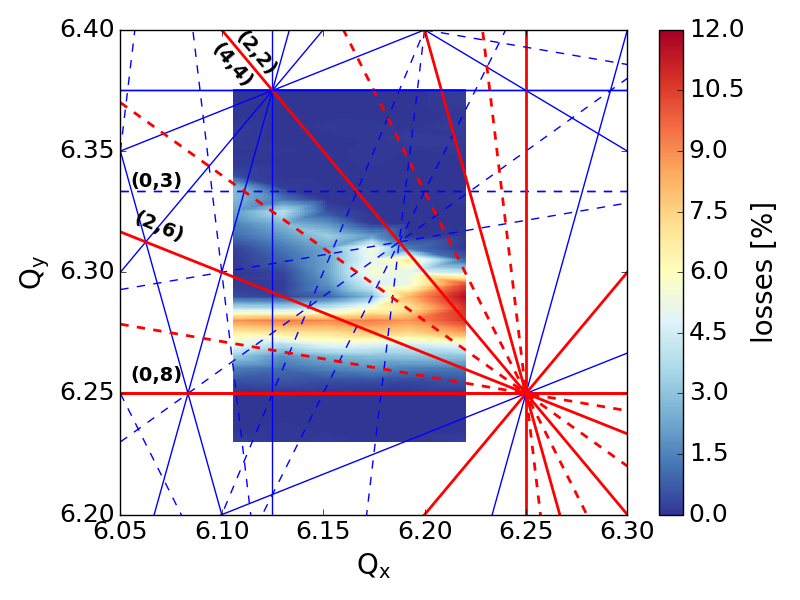}
        \includegraphics[width=0.494\textwidth]{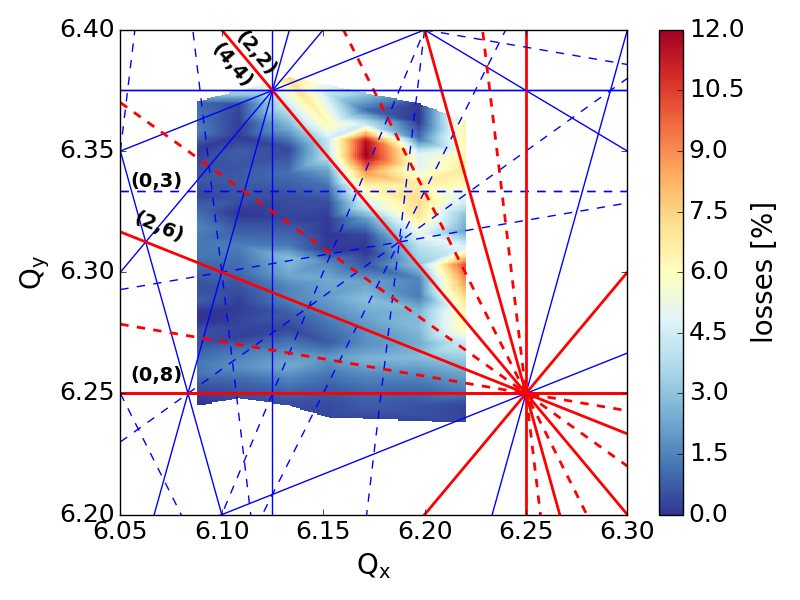}
        \includegraphics[width=0.494\textwidth]{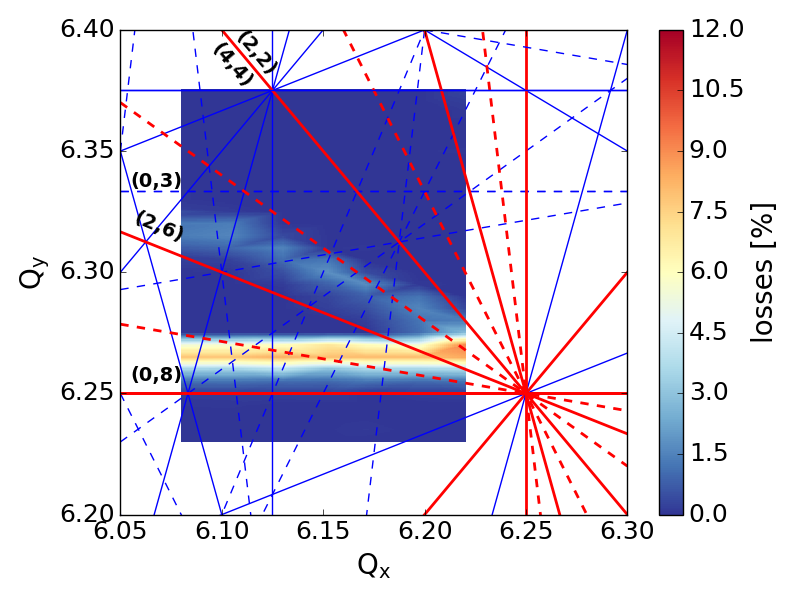}
    \caption{Tune scan with the $3Q_y=19$ resonance compensated. Results of the high-brightness (top) and medium-brightness (bottom) beams in measurements (left) and simulations (right). The color code represents losses after \SI{1.1}{s} of beam storage. Resonances of $\mathrm{3}^{\mathrm{rd}}$ and $\mathrm{4}^{\mathrm{th}}$ order are plotted, systematic in red and non-systematic in blue. The skew resonances are shown with dashed lines and the normal ones with solid lines. The resonances of interest are denoted as $(m,n)$ corresponding to the resonance condition $mQ_x+nQ_y=l$.}
    \label{fig:comp}
\end{figure*}

\begin{figure*}[!tb]
    \centering
   \includegraphics[width=0.494\textwidth]{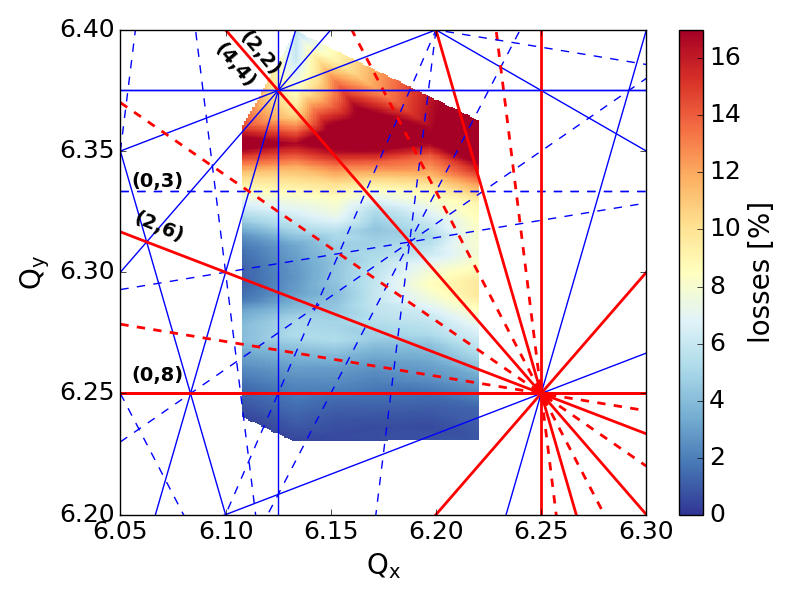}
   \includegraphics[width=0.494\textwidth]{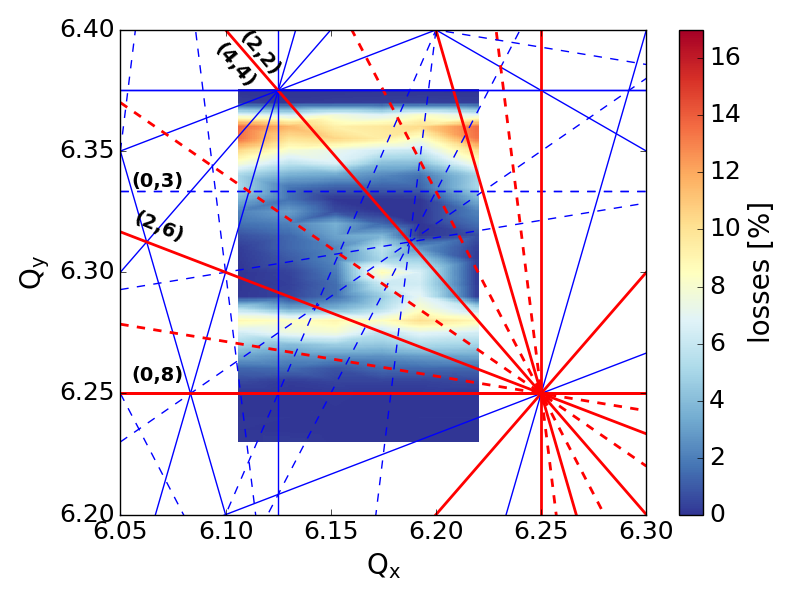}
   \includegraphics[width=0.494\textwidth]{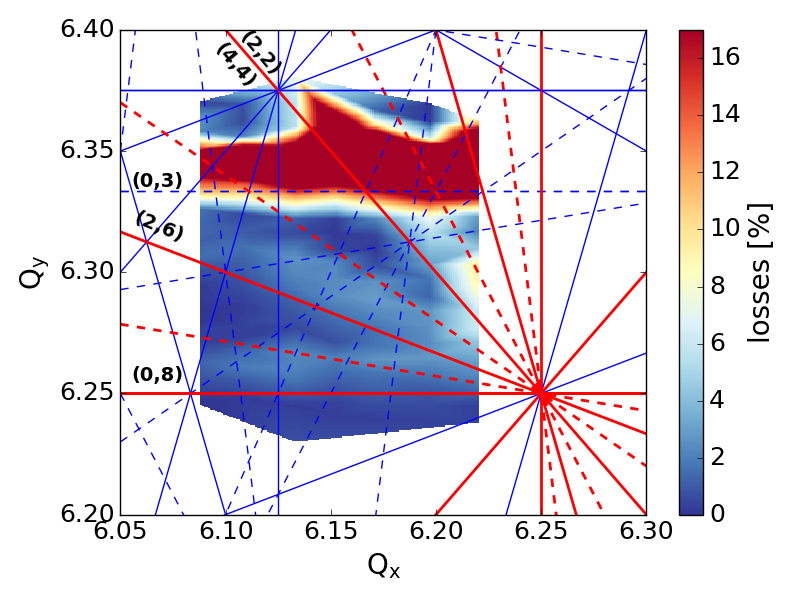}
   \includegraphics[width=0.494\textwidth]{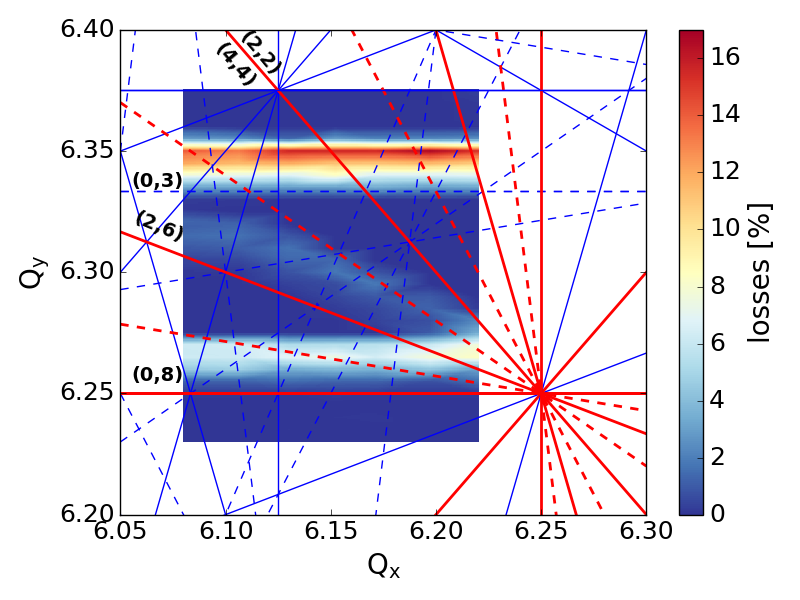}
    \caption{Tune scan with the resonance $3Q_y=19$ actively excited. Results of the high-brightness (top) and medium-brightness (bottom) beams in measurements (left) and simulations (right). The color code represents losses after \SI{1.1}{s} of beam storage. Resonances of $\mathrm{3}^{\mathrm{rd}}$ and $\mathrm{4}^{\mathrm{th}}$ order are plotted, systematic in red and non-systematic in blue. The resonances of interest are denoted as $(m,n)$ corresponding to the resonance condition $mQ_x+nQ_y=l$.}
    \label{fig:3}
\end{figure*}

Figure~\ref{fig:comp} shows the results of the static tune scans with the high-brightness and medium-brightness beams when the $3Q_y=19$ resonance was compensated.
The measured total losses over the $\SI{1.2}{s}$ long injection plateau are extracted for each measurement and the results are interpolated on a grid to identify the resonances, as shown in Fig~\ref{fig:comp} (left).
Likewise, each experimentally tested working point was simulated for $\mathrm{5\times10^{5}}$ turns, corresponding to the full PS injection plateau, using 3000 macroparticles with the adaptive frozen space charge solver in PyOrbit. Figure~\ref{fig:comp} (right) shows the resulting  losses interpolated on the same grid as before.
The PS model used in the simulations include systematic normal higher order field components obtained by matching the measured nonlinear chromaticity~\cite{Huschauer:1501943}.
Since the excitation of the resonance $3Q_y=19$ was compensated in the measurement, no extra skew sextupole-like components are needed in the PS model.

The measurements with the high-brightness beam show losses of about $7\%$ along tune values parallel to the $\mathrm{8^{th}}$ order resonance at 6.25.
Losses of similar magnitude are also observed along a line parallel to the coupled $\mathrm{8^{th}}$ order resonance at $2Q_x+6Q_y=50$.
Note that the offset between the resonance lines and the machine tunes at which losses are observed is a typical feature of the space charge induced periodic resonance crossing mechanism, as also discussed in Sec.~\ref{sec:lossmechanism}. 
The maximum losses are observed for large horizontal tunes $Q_x>6.18$ because the two resonances overlap and the bunch is affected by both of them at the same time.
In addition, losses of similar magnitude are observed in the vicinity of the anti-diagonal line. Since along this line losses are observed also with the low brightness beam during the dynamic tune scans discussed in Fig.~\ref{fig:alex}, and in addition the expected space charge induced RDT for the $4Q_x+4Q_y=50$ shown in Fig.~\ref{fig:RDTs} is weak, these losses are most likely caused by the $\mathrm{4^{th}}$ order resonance $2Q_x+2Q_y=25$ excited by octupole errors present in the ring.   

The measurements with the medium-brightness beam show losses along parallel lines to the space charge driven resonances $8Q_y=50$ and $2Q_x+6Q_y=50$, but in this case the observed losses of about $3-4\%$ are about half compared to the high-brightness beam, suggesting that the resonances are weaker due to the reduced space charge force.
On the other hand, losses on the resonance $4Q_x+4Q_y=50$ or rather $2Q_x+2Q_y=25$ appear instead enhanced. This further supports the excitation of the resonance in $\mathrm{4^{th}}$ order, since lower brightness beams are more sensitive to magnet error driven resonances.
Note also that the losses appear much closer to the respective resonances, since the incoherent space charge tune shift is smaller for the medium-brightness beam.

Comparing the experimental data with the corresponding simulations shows good agreement concerning the tune offset between the observed losses and the resonances causing them. Concerning the magnitude of losses, the space charge driven $2Q_x+6Q_y=50$ resonance seems to be very well reproduced both qualitatively and quantitatively. The quantitative agreement is not as good for the $8Q_y=50$ resonance, which results in slightly higher losses in the simulations compared to the measurements.
No losses are observed in the simulations along the $4Q_x+4Q_y=50$ resonance.
This is expected, since the analytic RDT calculation showed that the excitation of the $4Q_x+4Q_y=50$ resonance from space charge is weak, and no octupolar components are included in the model that could excite the $2Q_x+2Q_y=25$ resonance.
The fact that the modeling of the octupolar components appears to be incomplete could also affect the resonance at $Q_y=6.25$, contributing to the higher losses observed in the simulations compared to the measurements.
Presently, there are not enough independent octupole correctors available in the PS to compensate octupole-like resonances.
It should be noted that the resonance $2Q_x+6Q_y=50$ would remain unaffected by such components as in $\mathrm{4^{th}}$ order it would be $Q_x+3Q_y=25$ and thus a skew and not a normal resonance.

\subsection{Tune Scans with $\mathbf{3Q_y}$ Resonance Excited}

The results of the static tune scans with the high-brightness and medium-brightness beams in the presence of controlled skew sextupolar components are presented in Fig~\ref{fig:3}.
The controlled excitation of the $3Q_y$ resonance is achieved using a skew sextupole, the \enquote{XSK52}, powered at \SI{2}{A} corresponding to an integrated normalized skew sextupole strength of $k_{2\textrm{S}}l=\SI{0.0125}{\meter^{-2}}$.
Two more sextupoles, \enquote{XSK10} and \enquote{XSK14}, were always set to the compensation configuration described in Appendix~\ref{app:compensation}.
The data analysis and presentation is identical to the one described in the previous section and is given in Fig~\ref{fig:3} (left).
The simulations shown in Fig~\ref{fig:3} (right) are performed as discussed in the previous section, but in this case the sextupole \enquote{XSK52} was excited with the same strength as in the measurements.

The experimental data as well as the simulation results of both beams show the same behaviour as before for working points close to the space charge driven resonances, note the different scale of the color-code between Figs.~\ref{fig:comp} and~\ref{fig:3}, since their excitation is not affected by the addition of skew sextupolar components. 
Furthermore, the losses related to the $\mathrm{4^{th}}$ order $2Q_x+2Q_y=25$ observed in the experiment are also not affected by the nonlinearities introduced by the skew sextupole and as expected the simulations also show no response on this resonance.
In addition, the offset between the resonances and the tunes at which the losses are observed remain unaffected as no extra tune shifts are introduced.

The chosen excitation of the $3Q_y=19$ resonance results in much higher losses compared to the space charge driven structure resonances. It is worth pointing out that the relative losses from the  $3Q_y=19$ resonance are higher for the medium-brightness beam compared to the high-brightness beam, while the space charge driven resonances $8Q_y=50$ and $2Q_x+6Q_y=50$ resulted in higher losses for the high-brightness beam. This is another confirmation for the fact that the strength of the structure resonances depends directly on the space charge potential and is therefore proportional to the beam brightness.

\section{Conclusion}
\label{sec:conclusions}

Structure resonances driven by the space charge potential and the periodicity of the lattice were identified in the CERN PS.
By means of measurements at different integer tunes the dependency on the resonance harmonic and thus the lattice periodicity was verified.
Simulation studies of a simple FODO structure and the PS lattice itself demonstrated the incoherent nature of the high order space charge driven resonances and the periodic resonance crossing as the dominant loss mechanism.
Furthermore, the resonance driving terms of the space charge potential, calculated for resonances in the vicinity of the operational working point of the PS, showed the excitation of multiple $\mathrm{8^{th}}$ order resonances.
The presence of the $8Q_y=50$ and $2Q_x+6Q_y=50$ resonances was confirmed in a detailed experimental campaign, where it was demonstrated that their strength depends on the strength of the space charge potential and thus on the beam brightness. 
Moreover, the difference in the behaviour of the losses from the space charge compared to the lattice driven resonances further confirms the identification of the $8^\textrm{th}$ order incoherent space charge driven structure resonances in the PS.
The experiment could be reproduced using the simulation model of the PS. 
There are some slight differences between the measurements and the simulations, in particular concerning the beam loss in the vicinity of the anti-diagonal line not reproduced in the simulation, and the less good quantitative agreement for the $8Q_y=50$ resonance. 
This is, however, most likely attributed to the incomplete model of octupole-like errors in the PS lattice as the beam loss at those resonances could not be reproduced with the medium-brightness beam either.
The fact that the resonances in the vicinity of the nominal working point of the PS are space charge driven, justifies the higher injection energy in the scope of the LIU project as a means to accommodate higher brightness beams.

\begin{acknowledgments}

The authors would like to thank the operations teams of the PS complex, as well as H.~Damerau and G.~Sterbini for their help in the experiments, F.~Chapuis for the preparation of the beams, M.~Kaitatzi for providing the data of the dynamic tune scan, K.~Hizanidis for his help and support throughout the studies and G.~Franchetti, I.~Hofmann, J.~Qiang and F.~Schmidt for fruitful discussions. 

\end{acknowledgments}


\appendix

\section{Resonance compensation}
\label{app:compensation}

\begin{figure}[!b]
    \centering
     \includegraphics[width=0.7\columnwidth]{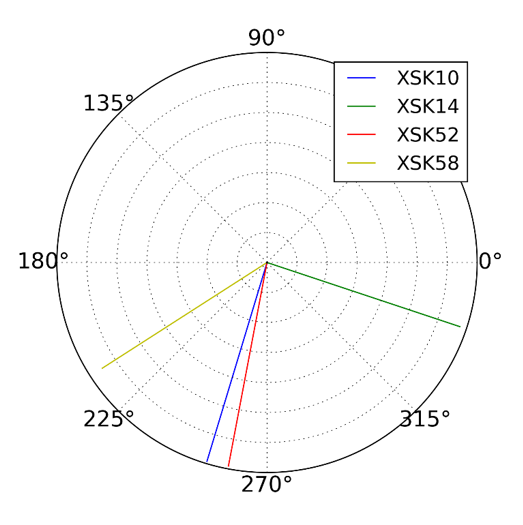}
     \includegraphics[trim=10 0 20 0, clip, width=0.86\columnwidth]{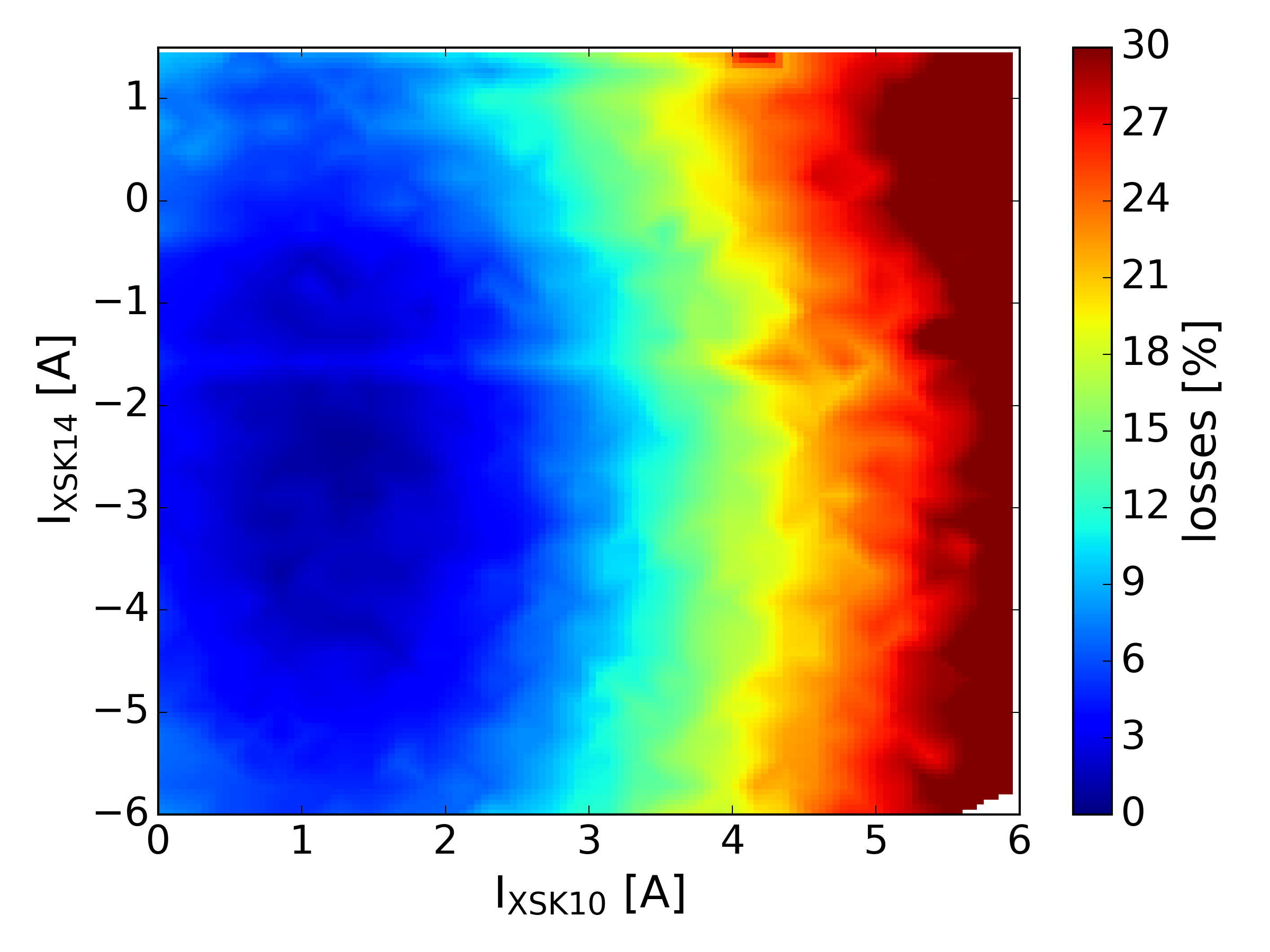}
     \includegraphics[width=0.86\columnwidth]{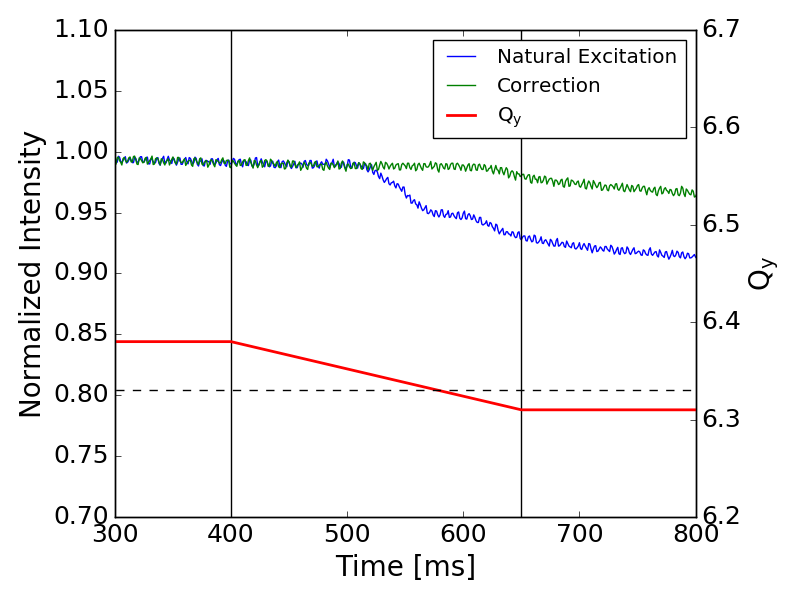}
    \caption{Comparison of the orientation and relative amplitude of the $3Q_y=19$ RDT generated by the skew sextupole correctors of the PS at same excitation (top). Scan of the currents of the \enquote{XSK14} and \enquote{XSK10} sextupoles with losses indicated by the color bar (centre), where $\SI{1}{A}$ corresponds to an integrated normalized skew sextupole strength of $k_{2\textrm{S}}l=\SI{6.25e-3}{\meter^{-2}}$.
	Normalized intensity along the injection plateau for the natural excitation and the correcting configuration of the sextupoles with the tune evolution indicated on a second axis (bottom).}
	\vspace{-0.7cm}
  \label{l2ea4-f2}
\end{figure}

The $\mathrm{3^{rd}}$ order skew resonance at $Q_y=6.33$, which was used with controlled excitation for the experiment, is naturally excited in the PS as shown in Fig.~\ref{fig:alex}. However, this resonance can be compensated using the available skew sextupole correctors. 
Figure~\ref{l2ea4-f2} (top) shows the strength and phase of the corresponding RDTs for the four available skew sextupole magnets when powered individually with the same current.
The \enquote{XSK14} and \enquote{XSK10} sextupoles were selected for compensating the resonance, since their RDT vectors are almost orthogonal.
Therefore, the full RDT space is accessible by powering them at different strengths in order to determine the best setting to compensate the unknown skew sextupole component in the lattice.
The described technique was applied using the medium-brightness beam so that any effects connected to space charge would be negligible. 
The beam was injected at $Q_{\mathrm{y}}=6.38$ and the (0,3) resonance was dynamically crossed from \SIrange{400}{650}{\milli\second}, as the tune was changed to $Q_{\mathrm{y}}=6.31$, while the currents of the selected skew sextupoles were varied shot after shot.
The losses as a function of the current configuration of the two sextupoles are shown in Fig.~\ref{l2ea4-f2} (centre). 
The sextupole XSK10 seems to be more effective, however, the magnets are identical so this difference is only correlated to the unknown phase and amplitude of the excitation and the locations of the sextupoles and the errors.
The configuration giving the least amount of losses was used for the correction, i.e.~powering XSK10 with $\SI{1}{A}$ and XSK14 with $\SI{-3}{A}$.
The effectiveness of the the compensation with the configuration chosen is demonstrated by the change of the slope in the intensity from \SIrange{400}{650}{\milli\second} for the natural excitation and after the correction, as shown in Fig.~\ref{l2ea4-f2} (bottom).

\section{Calculation of resonance driving terms from space charge potential}
\label{app:RDTs}

The RDTs are derived using the description and code in~\cite{Asvesta:2696190}.
This work is a generalization of the Hamiltonian description used in~\cite{Guignard:695942,MACHIDA1997316,Lee_2006} for any ring and any space charge RDT or detuning term.
In~\cite{Guignard:695942} the Hamiltonian description is used for the derivation of RDTs of any resonance, while the works in~\cite{MACHIDA1997316} and~\cite{Lee_2006} study $\mathrm{4^{th}}$ order space charge driven resonances.
Reference \cite{MACHIDA1997316} refers to the calculation of the resonance width of the $4Q_x$ space charge driven resonance in the KEK Proton Synchrotron, and in \cite{Lee_2006} the RDTs of all $\mathrm{4^{th}}$ order space charge driven resonances in a FODO lattice with a periodicity of 24 are calculated.
A short description of the derivation follows.

The integrand of Eq.~\eqref{eq:SCpotential} can be expanded, in both $x$ and $y$, using Taylor series under the paraxial approximation and the integral can be evaluated analytically.
The analysis of this paper focuses only on the $\mathrm{8^{th}}$ order of the potential, which yields:
\begin{equation*}
    \mathcal{V}_\textrm{sc}^{(8)}=K_{sc}(\Tilde{V}_{sc}^{(0,8)}+\Tilde{V}_{sc}^{(2,6)}+\Tilde{V}_{sc}^{(4,4)}+\Tilde{V}_{sc}^{(6,2)}+\Tilde{V}_{sc}^{(8,0)}),
\end{equation*}
\noindent
where $K_{sc}=\frac{r_0N_b}{\beta^2\gamma^3\sqrt{2\pi}\sigma_s}$, $r_0$ is the classical particle radius, $N_b$ the bunch intensity, $\mathrm{\beta, \gamma}$, the relativistic factors, $\sigma_s$ the longitudinal beam size and $\Tilde{V}_\textrm{sc}^{(m,n)}$ the part of the potential of the order $(m,n)$ in $(x,y)$, which reads:

\begingroup
\allowdisplaybreaks
\begin{align*}
    &\Tilde{V}_\textrm{sc}^{(0,8)}= \frac{(5\sigma_x^3+20\sigma_x^2\sigma_y+29\sigma_x\sigma_y^2+16\sigma_y^3)\cdot y^8}{6720\sigma_y^7\cdot(\sigma_x^4 + 4\sigma_x^3\sigma_y + 6\sigma_x^2\sigma_y^2 + 4\sigma_x\sigma_y^3 +\sigma_y^4)},\\
    &\Tilde{V}_\textrm{sc}^{(2,6)}= \frac{(\sigma_x^2+4\sigma_x\sigma_y+5\sigma_y^2)\cdot x^2\cdot y^6}{240\sigma_x\sigma_y^5\cdot(\sigma_x^4 + 4\sigma_x^3\sigma_y + 6\sigma_x^2\sigma_y^2 + 4\sigma_x\sigma_y^3 +\sigma_y^4)},\\
    &\Tilde{V}_\textrm{sc}^{(4,4)}= \frac{(\sigma_x^2+4\sigma_x\sigma_y+\sigma_y^2)\cdot x^4\cdot y^4}{96\sigma_x^3\sigma_y^3\cdot(\sigma_x^4 + 4\sigma_x^3\sigma_y + 6\sigma_x^2\sigma_y^2 + 4\sigma_x\sigma_y^3 +\sigma_y^4)},\\
    &\Tilde{V}_\textrm{sc}^{(6,2)}=  \frac{(5\sigma_x^2+4\sigma_x\sigma_y+\sigma_y^2)\cdot x^6\cdot y^2}{240\sigma_x^5\sigma_y\cdot(\sigma_x^4 + 4\sigma_x^3\sigma_y + 6\sigma_x^2\sigma_y^2 + 4\sigma_x\sigma_y^3 +\sigma_y^4)},\\
    &\Tilde{V}_\textrm{sc}^{(8,0)}=  \frac{(16\sigma_x^3+29\sigma_x^2\sigma_y+20\sigma_x\sigma_y^2+5\sigma_y^3)\cdot x^8}{6720\sigma_x^7\cdot(\sigma_x^4 + 4\sigma_x^3\sigma_y + 6\sigma_x^2\sigma_y^2 + 4\sigma_x\sigma_y^3 +\sigma_y^4)},
\end{align*}
\endgroup
\noindent
where $\sigma_{x}$ and $\sigma_{y}$ are the horizontal and vertical beam sizes, respectively.

Each of the above components of the field is Floquet transformed using 
$x(,y)=\sqrt{2\beta_{x(,y)} J_{x(,y)}}\cos{\psi_{x(,y)}}$
and gives resonance and nonlinear detuning terms~\cite{Lee:1425444}.
For instance the potential $\Tilde{V}_\textrm{sc}^{(0,8)}$ yields the terms:
\begin{widetext}
\begin{equation*}
    \Bar{V}_\textrm{sc}^{(0,8)}=\frac{5\sigma_x^3+20\sigma_x^2\sigma_y+29\sigma_x\sigma_y^2+16\sigma_y^3}{6720\sigma_y^7(\sigma_x^4 + 4\sigma_x^3\sigma_y + 6\sigma_x^2\sigma_y^2 + 4\sigma_x\sigma_y^3 +\sigma_y^4)}\cdot(\beta_yJ_y)^4\frac{1}{8}\cdot(56\cos{2\psi_y}+28\cos{4\psi_y}+8\cos{6\psi_y}+\cos{8\psi_y}+35),
\end{equation*}
\end{widetext}
\noindent
where $\psi_y = \Phi_y + \phi_y - Q_y\frac{2pi}{C}$, $(J_y,\Phi_y)$ are the action angle conjugate coordinates, $\phi_y$ the phase advance, $C$ the circumference of the ring and $\beta_y$ the vertical beta function. 

The angle independent term of the potential yields the nonlinear detuning, while the angle dependent terms drive resonances. These terms can be included as perturbations in the nonlinear Hamiltonian and the RDTs of $\mathrm{8^{th}}$ order resonances conjugate to the action are given as:
\begin{widetext}
\begingroup
\centering
\allowdisplaybreaks
\begin{align*}\label{eq:rdt}
    G_{(0,8)}&=\frac{K_\textrm{sc}}{16\pi}\int_0^{C}\frac{\beta_y^4\cdot(5\sigma_x^3+20\sigma_x^2\sigma_y+29\sigma_x\sigma_y^2+16\sigma_y^3)}{6720\sigma_y^7(\sigma_x^4 + 4\sigma_x^3\sigma_y + 6\sigma_x^2\sigma_y^2 + 4\sigma_x\sigma_y^3 +\sigma_y^4)}\cdot e^{j(8\phi_y)}ds,\\
    G_{(2,6)}&=\frac{K_\textrm{sc}}{8\pi}\int_0^{C}\frac{\beta_x\cdot\beta_y^3\cdot(\sigma_x^2+4\sigma_x\sigma_y+5\sigma_y^2)}{240\sigma_x\sigma_y^5\cdot(\sigma_x^4 + 4\sigma_x^3\sigma_y + 6\sigma_x^2\sigma_y^2 + 4\sigma_x\sigma_y^3 +\sigma_y^4)}\cdot e^{j(2\phi_x+6\phi_y)}ds,\\
    G_{(4,4)}&=\frac{K_\textrm{sc}}{16\pi}\int_0^{C}\frac{\beta_x^2\cdot\beta_y^2\cdot(\sigma_x^2+4\sigma_x\sigma_y+\sigma_y^2)}{96\sigma_x^3\sigma_y^3\cdot(\sigma_x^4 + 4\sigma_x^3\sigma_y + 6\sigma_x^2\sigma_y^2 + 4\sigma_x\sigma_y^3 +\sigma_y^4)}\cdot e^{j(4\phi_x+4\phi_y)}ds,\\
    G_{(6,2)}&=\frac{K_\textrm{sc}}{8\pi}\int_0^{C}\frac{\beta_x^3\cdot\beta_y\cdot(5\sigma_x^2+4\sigma_x\sigma_y+\sigma_y^2)}{240\sigma_x^5\sigma_y\cdot(\sigma_x^4 + 4\sigma_x^3\sigma_y + 6\sigma_x^2\sigma_y^2 + 4\sigma_x\sigma_y^3 +\sigma_y^4)}\cdot e^{j(6\phi_x+2\phi_y)}ds,\\
    G_{(8,0)}&=\frac{K_\textrm{sc}}{16\pi}\int_0^{C}\frac{\beta_x^4\cdot(16\sigma_x^3+29\sigma_x^2\sigma_y+20\sigma_x\sigma_y^2+5\sigma_y^3)}{6720\sigma_x^7\cdot(\sigma_x^4 + 4\sigma_x^3\sigma_y + 6\sigma_x^2\sigma_y^2 + 4\sigma_x\sigma_y^3 +\sigma_y^4)}\cdot e^{j(8\phi_x)}ds,\\
    G_{(4,-4)}&=\frac{K_\textrm{sc}}{16\pi}\int_0^{C}\frac{\beta_x^2\cdot\beta_y^2\cdot(\sigma_x^2+4\sigma_x\sigma_y+\sigma_y^2)}{96\sigma_x^3\sigma_y^3\cdot(\sigma_x^4 + 4\sigma_x^3\sigma_y + 6\sigma_x^2\sigma_y^2 + 4\sigma_x\sigma_y^3 +\sigma_y^4)}\cdot e^{j(4\phi_x-4\phi_y)}ds.\\
\end{align*}
\endgroup
\end{widetext}




\end{document}